\documentclass[oneside, a4paper, onecolumn, 11pt]{article}
\usepackage{times}
\usepackage[left=2cm,top=2cm,bottom=2cm,right=2cm]{geometry}
\usepackage[onehalfspacing]{setspace}
\usepackage{graphicx} % Required for inserting images
\usepackage{natbib}
\usepackage{tabularx}
\usepackage{booktabs}
\usepackage{makecell}
  \usepackage{threeparttable}
\usepackage{multicol}
\usepackage{pdflscape} % For landscape pages

\usepackage{amsmath,amssymb}
\usepackage[table]{xcolor} % Add this to use rowcolors
\usepackage{amsthm}
\newtheoremstyle{rpstyle}
  {6pt}{6pt}       % space above/below
  {\itshape}       % body font (ITALIC BODY TEXT)
  {}               % indent amount
  {\bfseries}      % head font (bold)
  {}               % punctuation after label
  {.25em}            % space after label
  {\thmname{#1}\thmnumber{#2}\thmnote{ (#3)}:} % header format: RQ1 (text)

\theoremstyle{rpstyle}
\newtheorem{researchquestion}{RQ}
\newtheorem{hypothesis}{H}

\setcitestyle{authoryear,open={(},close={)}}

\title{The Changing Global Division of Labor in Software: Emergence and Diffusion of New Programming Skills across IT Hubs
}

\author{Johannes Wachs$^{1,2,3,*}$, Xiangnan Feng$^{3}$, Simone Daniotti$^{3,4}$, Frank Neffke$^{3,5}$\\
\footnotesize{$^{1}$ Center for Collective Learning, Corvinus University of Budapest, Hungary}\\
\footnotesize{$^{2}$ Centre for Economic and Regional Studies, Hungary}\\
\footnotesize{$^{3}$ Complexity Science Hub, Vienna, Austria}\\
\footnotesize{$^{4}$ Copernicus Institute of Sustainable Development, Utrecht University, Utrecht, Netherlands}\\
\footnotesize{$^{5}$ Transforming Economies Lab, Interdisciplinary Transformation University Austria (IT:U), Linz, Austria}\\
\footnotesize{$^{*}$ Corresponding author: johannes.wachs@uni-corvinus.hu}
}
\date{}

\begin{document}

\maketitle
\begin{abstract}
With the rise of new industries, often new jobs emerge. Evolutionary Economic Geography and in particular Industry Life Cycle perspectives predict that these activities first emerge in a limited number of cities to then diffuse to other locations as job descriptions become more standardized. Here, we focus on a particularly important new industry: software development, an activity that is economically important, quickly changing, and has a  pronounced spatial concentration in a small number of global IT hubs. We use an online database of over 60 million questions and answers about problems in software development that yields a longitudinal dataset of 237 software skills. By geo-locating 3 million posting users at regular intervals, we link these skills to cities worldwide. We find that, in spite of its digital nature, the software industry exhibits similar  spatial regularities as previously observed in more traditional sectors. First, cities diversify into skills that are related to their existing ones. Second, new skills first emerge in cities with large and diversified software sectors, and later diffuse ---mostly unhindered by geographical distance--- to smaller cities specialized in closely related skills. We find suggestive but limited support for a windows of locational opportunity account: although even brand-new skills still emerge first in cities with strong prior specialization in related skills, concentrations of related activities impact less the emergence of new skills than the diffusion of existing ones.
\end{abstract}

\bigskip \parbox[t]{.95\linewidth}{\textbf{Keywords} $ $ skills $\cdot$ software sectors $\cdot$ economic geography $\cdot$ industry life cycle $\cdot$ technological diffusion $\cdot$ relatedness}

\section{Introduction}
Software development has quickly become one of the 21st century's most important industries. Eight out of ten of the largest companies worldwide are either heavily engaged in development of software or providing the hardware and infrastructure that software runs on. On the labor market, advances in software in the broad sense have raised anxieties about an uncertain future of work. At the same time, the number of workers employed in jobs that require familiarity with coding has grown fast. Unsurprisingly, countries and cities around the globe have started an intense competition for dominance of, or at least participation in, the frontier of software development. With the recent advance in large language models in particular and AI in general, this competition has intensified to the level where it is having geopolitical consequences. Yet we know remarkably little about the changing nature of software development jobs or how the global division of labor in this sector has evolved. 

Here, we aim to make progress by leveraging a large dataset that documents community-generated problems and solutions in programming that can be used to create a detailed taxonomy of software skills \citep{feng2025building}. We study these skills through the lens of key conceptual frameworks in evolutionary economic geography (EEG). This allows us to gain insight into how cities diversify within the software development sector, as well as where new programming skills first emerge and  then diffuse across the globe.   

To analyze the software sector, we take inspiration from four conceptual frameworks in EEG. First, building on the literature on economic complexity, and in particular the strand of this literature that has analyzed the nature of jobs and human capital \citep{alabdulkareem2018unpacking,anderson2017skill,neffke2019value}, we aim to understand software jobs at the level of the skills that programmers use at work. Second, following the literature on the principle of relatedness \citep{hidalgo2018principle}, we construct networks of related skills to describe the software skill bases of cities, mimicking work that uses industrial and technological relatedness matrices  \citep[e.g.,][]{neffke2011regions,boschma2015relatedness,kogler2015mapping}. This allows us to rank cities in terms of the economic value of their software skill base and study the importance of relatedness in the evolution of a city's local software sector. Third, we test whether the windows of locational opportunities concept \citep{boschma1997new,walker1989capitalist} helps understand the geography of new skills. Fourth, and most importantly, we take inspiration from the product \citep[PLC, ][]{vernon1966international} and industry life cycle literature \citep[ILC,][]{klepper1997industry} to analyze the evolution of the software labor market. According to geographical interpretations of product and industry life cycle theories, new industries typically emerge in large, diversified cities and later diffuse to smaller, more specialized and peripheral locations \citep{duranton2001nursery,neffke2011dynamics}. Most of the EEG literature on ILCs has focused on the evolution of agglomeration externalities and industrial clusters, studying, for instance, spin-off dynamics and firm survival rates \citep{klepper2007disagreements,boschma2007spatial}. Yet, an arguably equally important, yet often overlooked, aspect of ILCs is how progressive technological standardization over the course of an industry's life cycle impacts skill requirements and careers.  

Lack of longitudinal, high-resolution data on changing job requirements along the ILC has so far made it hard to study where new  skills emerge and how they diffuse across locations. To remedy this, we split software development expertise into fine-grained skills. We do so by exploiting repeated data dumps from Stack Overflow (SO), an online knowledge repository for software developers. The data are structured in a question-answer format, covering over 60 million questions and answers and 10 million users across hundreds of computer languages. Users can post questions about a specific software problem they try to solve. Other users can reply to these posts by offering solutions. For retrieval purposes, questions are tagged with a system of tens of thousands of distinct labels.  Following \cite{feng2025building}, who construct a taxonomy of software development skills based on the  tags in SO, we interpret tags as structured descriptions of the software challenges described in each question. Next, these authors run advanced community detection algorithms to identify clusters of tags that often co-appear on the same questions. These clusters represent  generalized recurring problems in software development. Similarly, users' ability to  answer the associated questions indicates their mastery of the skills needed to tackle such classes of problems. We therefore refer to these tag clusters as ``skills'' and label them with concise natural-language descriptions to arrive at a dataset of software development skills.

Having defined skills as groups of tags, we analyze the co-occurrences of these skills in the posting history of users on SO. That is, we assume that providing answers (not merely asking questions) signals relevant skills and thus connect users to skills that are associated with the questions they answer. Following \cite{anderson2017skill} and \cite{alabdulkareem2018unpacking}, we measure how related different skills are to one another based on how often they are practiced by the same users. This leads to a network, the \emph{software skill space}, that describes the relations among software skills. 

Despite the digital nature of software production, traditional EEG theories  prove to offer a powerful framework for understanding the spatial dynamics of this sector. Our findings demonstrate that nearly all novel skills introduced in software development work since 2010 first appear in a remarkably small number of large and diversified cities, including Silicon Valley, Seattle and London. Moreover, cities diversify preferentially into software skills that are closely related to the ones they already master. Finally, geographic distance beyond the city boundary only weakly constrains the diffusion of new ideas in this sector. In line with this, we see suggestive evidence that windows of locational opportunity have opened up: having related skills matters less for the emergence of new skills than for the diffusion of existing ones, though new skills still tend to emerge first in cities with related prior expertise.

Our work contributes to a number of different debates. First, it connects to the  literature in economic complexity  on path dependent  development \citep{frenken2007theoretical,martin2006path} and the principle of relatedness \citep{hidalgo2018principle,neffke2011regions}. Second, we connect to a rapidly growing literature that uses high-dimensional skill and task vectors to describe human capital \citep{alabdulkareem2018unpacking,anderson2017skill,neffke2019value,stephany2024price,hosseinioun2025skill}. We add to this body of research by  zooming in on the world of software development. This allows us to contribute to the debates on the future of work in a  novel way by studying how programming skills diffuse across cities. This is important, given that programming skills are becoming increasingly demanded across a growing number of jobs \citep{brynjolfsson2014second}. Third, we contribute to the literature in economic geography on agglomeration externalities \citep{marshall1920industrial,rosenthal2004evidence,duranton2004micro}, clusters \citep{porter2003economic}, and their life cycles \citep{menzel2010cluster,neffke2011dynamics} and especially of IT clusters \citep{saxenian1996regional,chattergoon2022winner}. Finally, our approach allows mapping the software sector ---nowadays the dominant component of most of these clusters--- and its spatial footprint in unprecedented detail. Moreover, software is integrated into the production processes and final output of an increasing number of industries. Our approach and datasets open up the possibility to analyze the interaction between software and traditional industries at a fine-grained geographic detail. To facilitate such future research, we make the detailed software-development skill profiles we generate available for functional urban areas around the world.

The paper is organized as follows. Section~\ref{sec:litrev} describes relevant prior literature and uses this to  formulate  research questions and derive hypotheses. Section~\ref{section2data} describes the dataset. Section~\ref{section3methodology} describes how we observe skills and link them to cities. Section~\ref{section4results} presents results on the economic value of software skills, the principle of relatedness in software development, and the geography of the emergence and diffusion of new software skills. Section~\ref{section5discussion} summarizes the work.

\section{Literature and research questions}\label{sec:litrev}

\subsection{The evolution of software development}

Software has been around since at least 1948, when Williams and Kilburn ran the first stored program on the Manchester ``Baby'' machine. Yet software as a commercial product is younger. Until 1969, software was bundled with hardware: IBM gave away programs to sell machines. When IBM ``unbundled'' software from hardware under antitrust pressure, it created the conditions for an independent software industry \citep{campbell2003history}. A new professional discipline coalesced around the NATO Software Engineering conferences of 1968-69, which coined the term ``software crisis'' to describe the growing gap between what hardware could do and what programmers could reliably build \citep{ensmenger2010computer}. The personal computer revolution of the 1980s then transformed software from a bespoke service into a mass-market product. The commercialization of the internet in the mid-1990s brought another wave of change, giving rise to web applications, e-commerce, and the open-source movement \citep{lerner2002economics}. And since 2000, cloud computing, mobile platforms, and software-as-a-service have further expanded the sector's reach \citep{campbell2009saas}. 

Today, software pervades virtually every industry. For example, roughly 40\% of the cost of a new car is attributable to electronic systems, most of them software-controlled \citep{charette2021car}, and digital product exports are growing at roughly three times the rate of physical goods trade \citep{stojkoski2024estimating}. The software industry is both the origin of the recent revolutionary developments in AI, and its primary frontier \citep{daniotti2026using}.

% We need to show that although software is not new, there is a clear renewal of the sector in or right before the period we study  

\begin{figure}
    \centering
    \includegraphics[width=0.75\textwidth]{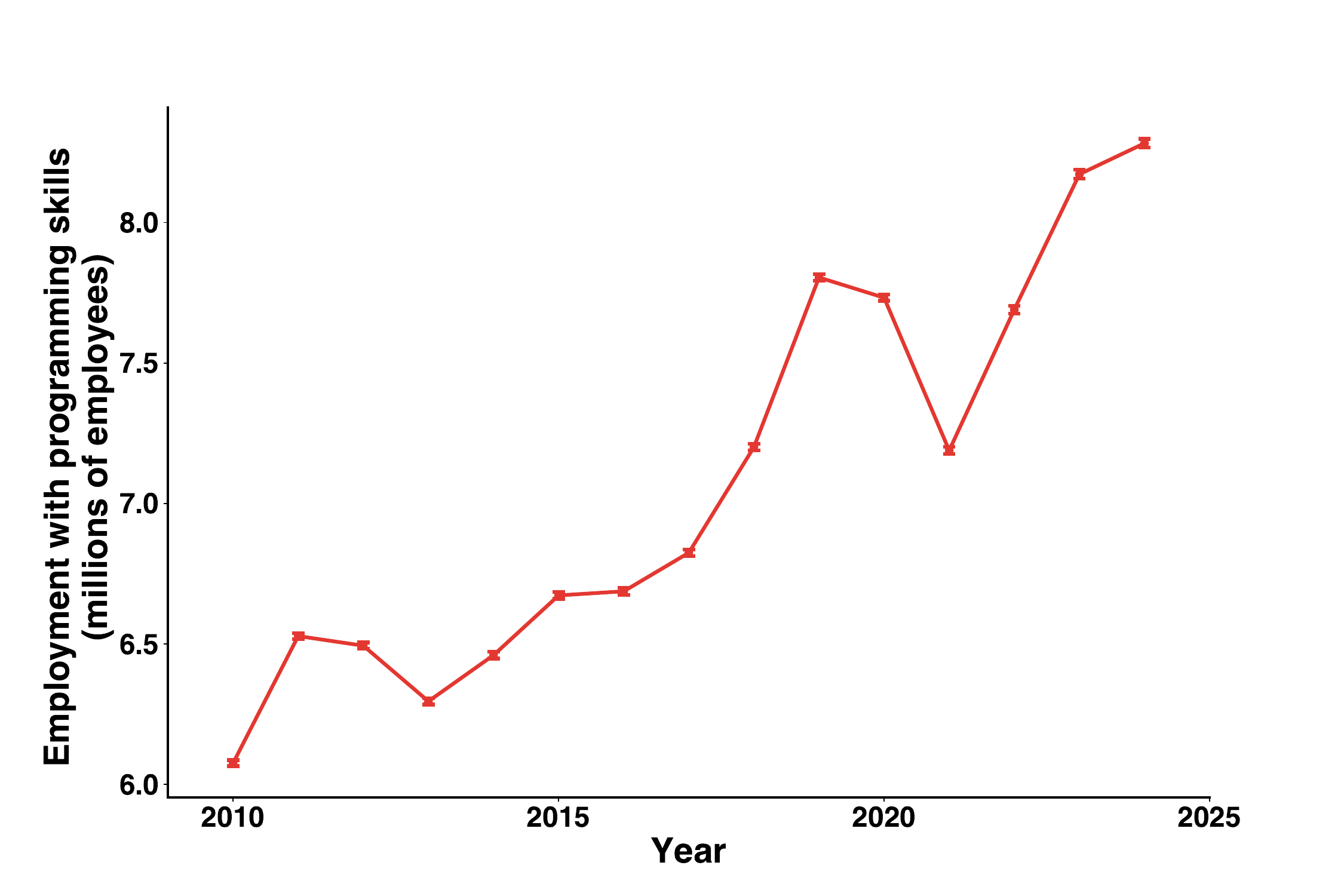}
\caption{U.S. employment in occupations requiring at least basic programming skills. Estimates are based on employment reported in the Occupational Employment and Wage Statistics from the U.S. Bureau of Labor Statistics in occupations that report requiring the skill ``Programming'' at level 2 (with anchor ``Write a program to sort objects in a database'') or higher.}
    \label{fig:employment}
\end{figure}

Software work is changing rapidly. According to our estimation, employees on job positions that need programming skills keep increasing, with the number of positions in the U.S. that require basic or higher programming ability surpassing 8 million in recent years (see Fig.~\ref{fig:employment}).
The skills required to produce software are also changing rapidly: in few other major industries are jobs and their constituent tasks changing as quickly as in software development. Beyond the computerization of routine tasks \citep{autor2003skill} and the proliferation of programmed robots \citep{graetz2018robots}, software skills are likely to be important complements in sophisticated economic activities, and therewith drivers of future economic growth \citep{bresnahan2002information}.

% geography of software work
Not only its dynamism sets software production apart. Software development also has a pronounced geography. In fact, software development presents an interesting puzzle for economic geographers. On the one hand, software is non-physical in nature. Consequently, software products can be traded over long distances at virtually no costs. Moreover, many firms outsource software activities across countries, either to access low wages or to shorten development cycles by taking advantage of time zone differences. On the other hand, software development is highly non-routine work requiring sophisticated know-how, which is often associated with a high degree of tacitness \citep{ryan2009development}, and collaboration \citep{faraj2000coordinating,demarco2013peopleware}. This would put a premium on geographic proximity. In line with this, previous work confirms that developers and their collaborations are intensely concentrated in space \citep{takhteyev2012coding,wachs2022geography,goldbeck2025bit}. Previous work at the national level \citep{korkmaz2024github,juhasz2026software} uses GitHub data to construct country-level measures of software production and complexity that predict economic outcomes, but the coarse geographic and content resolution of their analysis leaves open how software capabilities are distributed and evolve within countries.

\subsubsection{Evolutionary Economic Geography}
To make sense of the spatial characteristics and dynamics of the software sector, we turn to Evolutionary Economic Geography \citep[EEG,][]{boschma2006economic,boschma2007constructing,boschma2018evolutionary}, which offers a useful framework to analyze the changing geographical footprint of an industry. In particular, we build on three different strands in this literature: industry life cycle theory (ILC, \cite{klepper1997industry}), the literature on related diversification \citep{frenken2007theoretical,neffke2011regions,boschma2015relatedness,hidalgo2018principle}, and the windows of locational opportunity (WLO) concept \citep{boschma1997new}. 

ILC theory describes changes in an industry's innovation processes, competitive structure, and economic geography as its  core technologies mature. In an industry's initial stages, innovation tends to be radical and  change the qualitative nature of the new product or service. Such innovations often rely on Schumpeterian new combinations \citep{schumpeter1912theorie}, which are  supported by diverse urban economies acting as ``nurseries'' for new ideas \citep{jane1969economy,duranton2001nursery}. \cite{boschma1997new} proposes that in these initial stages, local factors and specializations are less critical, opening a window for a larger number of locations to act as incubators of such new industries.  As new industries mature, they transition from a stage of exploration to a stage of exploitation \citep{suarez1995dominant,klepper1997industry}. Innovation shifts from radical to incremental and from product to process innovation. The shake-out that forces the majority of firms out of the market leads to a consolidation during the exploitation  phase of an industry's life cycle. At this point, activity often concentrates in a few dominant clusters \citep{buenstorf2009heritage,klepper2007disagreements}. However, the emergence of a dominant design also evens the way for standardization, which in later phases allows the activity to diffuse to new, lower cost, locations \citep{vernon1966international,neffke2011dynamics}. 

One often hypothesized but rarely tested component of ILC theory is its predictions about changes in the labor force. For instance, ILC ---as well as the related literature on techno-economic paradigms \citep[e.g.,][]{freeman1988structural}--- proposes that changes in an industry's technology will also be reflected in the human capital it employs. In the initial stages of the life cycle, there is much freedom in how to design a product and uncertainty about the industry's future technological trajectory. Firms often experiment with a wide range of technologies, which requires a workforce that is flexible and highly skilled, but not overly specialized. Once a dominant design emerges, the shift towards incremental and process innovation leads to a more predictable technological trajectory and standardized production processes allow for a stable and deeper division of labor with high levels of specialization.  This standardization, in turn, allows companies to formalize and codify training, such that workforces can be hired and trained outside the initial hubs of the industry. Taking advantage of cheaper labor in more peripheral locations, the industry then diffuses spatially and across countries \citep{vernon1966international}.

\subsection{Economic complexity and skill spaces}
At the level of skills, the ILC framework suggests that new skills emerge in the industry's core clusters. However, WLO theory suggests that for more radical innovations opportunities for leapfrogging may exist in a wider set of regions. In either case, as new skills become better understood, jobs become more standardized and start diffusing out of the high-cost technology hubs where they originated. However, lack of longitudinal, high-resolution data on job tasks has so far made it hard to test this hypothesis. This is changing with what can be described as the emergence of an \emph{economic complexity approach to human capital}. This approach studies either the detailed movements of workers across jobs to infer which jobs are related in the sense that they require similar skills \citep[e.g.,][]{neffke2013skill,del2021occupational,o2022modular} or fine-grained information on the skills and tasks of workers  \citep{gathmann2010general,anderson2017skill,alabdulkareem2018unpacking,neffke2019value,stephany2024price,del2025enhancing}. 

The latter information is subsequently used to map abstract skill spaces that  express the similarity, synergy or complementarity between skills. Jobs can subsequently be projected onto these spaces, to highlight the skills that a given occupation requires. For instance, \cite{anderson2017skill} constructs skill spaces based on tags that describe workers and jobs on an online freelance platform, showing that wages depend on how well users are able to combine skills that are often demanded in the same jobs, but rarely provided by the same workers. Using similar data, \cite{stephany2024price} show that their complementarity to a wide variety of existing skills explains the value of Artificial Intelligence (AI) skills in freelance work. Recently, scholars have started looking at  regional implications \citep[e.g.,][]{frank2024network,henning2025job} finding that cities exhibit pronounced specialization patterns in such skill spaces that impact on workers careers.

\subsubsection{Research questions and hypotheses}
Because software is at its very core a digital activity --- with products that can be moved at zero transportation costs and workers who  collaborate on online collaboration platforms  untethered to specific locations --- it is unclear whether the canonical explanatory frameworks that have been used to describe spatial dynamics of traditional industries still apply. With respect to the product and industry life cycle theory, we therefore ask:

\begin{researchquestion}[relatedness]\label{rq:relatedness}
How do cities diversify their software skill portfolios?
\end{researchquestion}
\begin{researchquestion}[emergence]\label{rq:emergence}
Where do new skills in software development first emerge?
\end{researchquestion}
\begin{researchquestion}[diffusion]\label{rq:diffusion} 
Where do these skills subsequently diffuse to?
\end{researchquestion}

Our baseline hypothesis is that work in the software industry follows the general patterns predicted by standard EEG theories. Following the  related diversification literature, we hypothesize that, despite the digital nature of software production, diversification patterns follow the standard principle of relatedness \citep{hidalgo2018principle}: 

\begin{hypothesis}\label{hyp:PoR} 
Cities preferentially move into new skills that are closely related to the ones they are already active in.
\end{hypothesis}

Similarly, we hypothesize that the emergence  of new skills follows the predictions by \cite{jane1969economy} and \cite{duranton2001nursery} that novelty thrives in large cities where a large number of diverse ideas can easily come together. Acknowledging that software is quite distinct from most other economic activities in terms of the nature of technologies (algorithms), labor markets and international competition, we assume that the diversity of ideas that matter is better captured by the diversity within the software sector than the urban environment as a whole:

\begin{hypothesis}\label{hyp:emergence}
New skills emerge in large cities with  diversified software sectors.
\end{hypothesis}

Furthermore, in line with WLO arguments, we expect that local factors matter less for the emergence of new skills than for the diffusion of existing skills to new locations. 

\begin{hypothesis}\label{hyp:WLO}
Local specialization in related activities is less constraining for where new skills emerge than for where existing skills are adopted.
\end{hypothesis}

We furthermore propose that the standard PLC forces as described by \cite{vernon1966international} and the nursery cities model of \cite{duranton2001nursery} also explain the subsequent diffusion of new skills to secondary locations:

\begin{hypothesis}\label{hyp:diffusion} 
New skills diffuse preferentially to nearby cities and cities specialized in closely related skills.
\end{hypothesis}

That is, we hypothesize that also    the software sector is constrained by geography. The reason is that knowledge used in software production, like in any high-skill activity, will contain an important tacit  component. This anchors software activities in specific locations. However, we expect that these centripetal forces are particularly strong at short distances, where face-to-face contacts can be easily organized. At longer distances, programmers can rely on the codified knowledge available in documentation, existing computer code and platforms like SO \citep{goldbeck2025bit}. Moreover, communication in software is facilitated by the fact that many details that need to be shared between collaborators are embedded in code composed in computer languages, which unlike natural languages are shared across the world. Moreover, code is often annotated in English. This overcomes linguistic problems in collaboration such that common projects can be  organized with almost equal ease between countries as between cities. 

For these reasons, we expect that the hyper-local effects captured by the principle of relatedness --- which refer to dynamics that play out within the same city or region --- are stronger than the local diffusion effects of the PLC and ILC literature that occur between cities:

\begin{hypothesis}\label{hyp:horserace} 
After their first emergence in the places of origin,  the later adoption of new skills by other cities depends more on how related these skills are to the city's existing software activities than on how close the city is to the spatial origins of a skill.
\end{hypothesis}

\section{Data}
\label{section2data}

Our primary source of data to test these hypotheses is Stack Overflow, the largest online question and answer platform for computer programming and software development topics. Founded in 2008, Stack Overflow regularly provides dumps of data on all questions and answers (together referred to as posts) and posting users. Stack Overflow questions are supposed to ``focus on questions about an actual problem'', and users are prompted to include specific details and even example code\footnote{https://stackoverflow.com/tour}. A sophisticated system of automated bots and community volunteers checks new posts for both appropriateness and novelty, minimizing duplicate questions and spam in the data. 

A growing literature studies why people contribute to Stack Overflow or open source projects in general. Contributors are motivated by a combination of intrinsic and extrinsic factors. On the intrinsic side, programmers report intellectual
stimulation, learning, and a desire to help others as key drivers \citep{lakhani2005hackers}. On the extrinsic side, Stack Overflow's reputation system, which awards points and badges for useful contributions, creates strong incentives for sustained participation \citep{anderson2012discovering}. Crucially, \citet{xu2020makes} show that career concerns are a key driver: users reduce their reputation-generating activity by roughly 24\% after finding a new job, suggesting that many treat their Stack Overflow profile as a labor market signal. User profiles on the platform are indeed widely used in hiring \citep{marlow2013activity}. These motivations matter for our analysis because they imply that contributions reflect genuine technical expertise rather than noise: users answer questions in domains where they can credibly demonstrate competence, and the community's voting system surfaces the most reliable content \citep{mamykina2011design}.

The corpus we derive from Stack Overflow is a large collection of problem-centered posts. We will use Stack Overflow posts in a similar way as canonical sources of data used to study the dynamics and geography of innovation, such as patents \citep{mansfield1986patents,kogler2013mapping,balland2017geography}, trademarks \citep{castaldi2018trademark}, or research papers \citep{mansfield1991academic,guevara2016research,nagaraj2026geography}. Stack Overflow data offer several advantages over these more established indicators. First, scale: with over 24 million questions and 10 million registered users, SO rivals major patent databases in volume. Second, temporal resolution: posts are timestamped to the day, whereas patents involve multi-year lags between invention and grant. Third, geographic coverage: SO draws contributors from over 150 countries on any given weekday, avoiding the jurisdictional biases of patent offices that concentrate observations in the US, Europe, and Japan. Fourth, industrial scope: patents are disproportionately filed in manufacturing, leaving the service sector largely unrepresented \citep{dorner2018concordance,griliches1990patents}; SO, by contrast, captures software work across all sectors. These advantages come with a caveat: SO reveals what developers know how to do, not necessarily what they produce. In this sense, our data are closer to a map of skills than of output \citep{acemoglu2011skills}. Nonetheless, the low barrier to posting and the problem-centered nature of the content give us unusually detailed information about the work activities and challenges that programmers face across the world.

Besides the text content of each post, questions are annotated with tags, which describe relevant technologies, methods, and concepts. For instance, a question about an algorithm to sort a list of elements or array in the Python programming language may be tagged with ``algorithm'', ``sorting'', ``list'', ``array'', and ``Python''. Tags are curated and deduplicated by the community to facilitate search and filtering. Analysis of tag co-occurrence and usage patterns has been used to study the innovation dynamics of programming technologies, tracking the rise and fall of tools and frameworks as the field evolves \citep{borchers2025innovation}. We follow \citet{feng2025building} and use data on which tags co-occur within posts to define a generalized notion of specific programming problems or tasks. Next we interpret SO users' ability to answer (not only ask) questions in a specific domain as a sign that they have expertise (``skills'') in this domain.

\subsection{Users and geography}
Stack Overflow user profiles contain information about the user and their track record and accomplishments on the site.  Given their value as labor market signals, users often annotate their profiles with personal information, including their real names and workplaces. Profiles contain an optional text field for users to provide their locations. We use this information to infer where posters are located. As users can update their locations, we collected user data from past dumps, aiming to build a time series of user locations from 2010 to the present.

We geolocated the raw user-strings using the Microsoft Bing Maps API. This API accurately assigns city, region, and country values from raw input strings. It handles multiple languages, for example ``Vienna'', ``Wien'', and ``B\'ecs'' are all interpreted as Vienna, Austria. We use this geographic information to annotate user profiles with latitudes and longitudes. Whenever we can infer a user's city, we use the city's latitude and longitude data to map them to Functional Urban Areas (FUAs) \citep{moreno2021metropolitan}, yielding a sample of three million geocoded users. FUAs use satellite data to merge connected urban areas and commuting zones. They identify urban agglomerations that form coherent labor markets. This avoids artificially splitting up agglomerations whose administrative boundaries do not have a significant influence on economic activity. For instance, San Jose, Palo Alto, and San Francisco are merged into a single FUA. Whenever users only report a country (e.g., ``France'') we do not assign them to an FUA. Following this procedure, we are able to assign just over three million posting users to FUAs. For conversational convenience, we will hereafter use the terms ``FUA'' and ``city'' interchangeably in the paper.

\subsection{Survey}
Stack Overflow runs an annual user survey which attracts tens of thousands of respondents. Participants are asked about their coding abilities and experience, programming language and technology use, work and salary, and other topics. We convert salaries into annual US dollars including bonuses. We will use this information to get approximate estimates of the value of different software skills.   %Although we cannot link survey respondents (which are anonymous) to user profiles, we can link salaries to specific languages and technologies that respondents use. By next linking these languages and technologies to tags in the question-answer data we will be able to estimate the labor market value of software skills that cities specialize in to generate global rankings of cities in terms of the value of the software skills they perform.

\section{Methodology}
\label{section3methodology}

\subsection{Constructing a classification of software skills}
\begin{figure}
    \centering
    \includegraphics[width=0.9\textwidth]{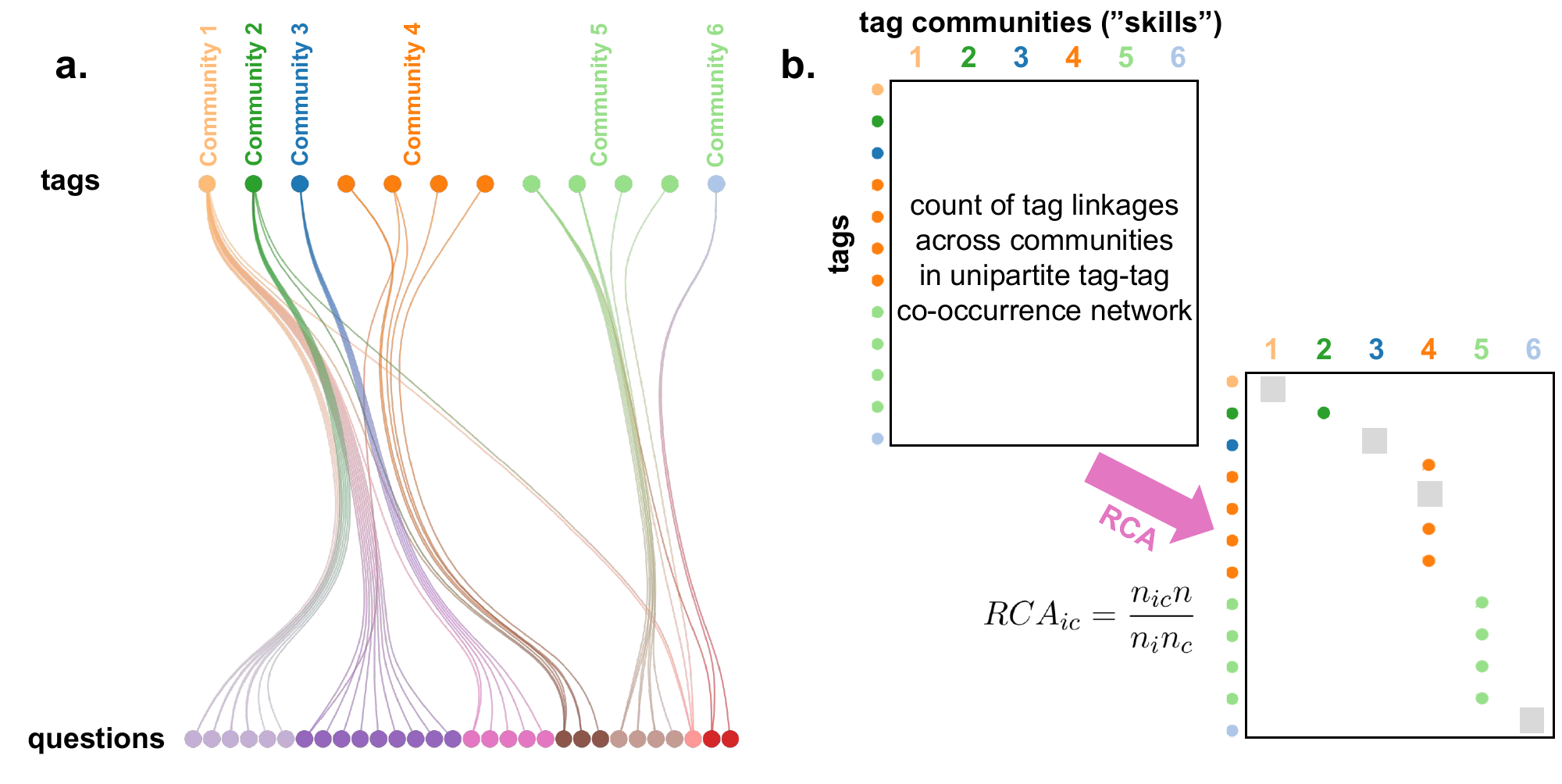}
\caption{{\bf a.} Stylized illustration of question-tag bipartite network. Different colors refer to different communities identified by the SBM. {\bf b.} Illustration of tag-community RCA calculations. After calculating the RCA, the bottom $20\%$ tags are removed, denoted by the gray box.}
    \label{fig:sbm}
\end{figure}

To extract generalized skills from Stack Overflow, we use the methodology developed in \citet{feng2025building}. In brief, these authors construct a bipartite network connecting 5,083 frequently used tags to over 18 million questions. They then apply a Stochastic Block Model \citep[SBM,][]{holland1983stochastic, peixoto2014hierarchical} to discover communities of tags that are linked to similar sets of questions (see Fig.~\ref{fig:sbm} for an illustration). Unlike standard community detection methods, SBMs cast this as a statistical inference problem, reducing the number of ad hoc modeling decisions and allowing for the quantification of uncertainty and confidence \citep{peixoto2023descriptive}. After removing weakly attached tags\footnote{The SBM can in principle result in assignments of nodes to communities where they maintain no direct co-occurrence links to any of the other nodes in the community. The assignment in such cases relies on indirect links. To tighten community definitions, \cite{feng2025building} first turn the bipartite question-tag network into a unipartite tag-tag network by counting the number of co-occurrences of tags in questions. Next, they count for each node in each community how connected they are to other nodes in the same community. They then determine whether within-community connections are overrepresented for this particular node, calculating the RCA of the number of links the node maintains with each community. Finally, the authors drop nodes that are in the bottom quintile in terms of this within-community linkage-RCA.} and small communities (fewer than 3 tags), this procedure yields 237 skill communities consisting of 4,054 core tags. \cite{feng2025building} interpret these communities as canonical software skills and label them with concise natural-language descriptions. For instance, one skill on natural language processing (NLP) combines \emph{spacy} (a Python library for NLP tasks), \emph{text-mining}, and \emph{word-embedding} (a central concept in large language models); others describe neural networks, mobile app development, database management, blockchain technologies, and so on. Full methodological details are provided in \cite{feng2025building}.

\subsection{Determining the composition of software skills in a city}
We can use the skills constructed above to describe the skill profiles of users in SO. To do so, we look at the tags associated with the questions to which a user posted answers. To be precise, following \cite{feng2025building}, we count the total number of answers a user provides on SO that engage in this way with each tag. This results in a 4,054-dimensional tag vector for each user. Next, we reduce the dimensionality of this vector by aggregating tags into skills, using the tag communities we constructed above. As a result, we can describe each user by a 237-dimensional skill vector. 

For most users, this skill vector is sparse, with many zeros and few relatively large non-zero values. Because each user only has a limited number of working hours available, we normalize these skill vectors to add up to 1. We will think of these numbers as an estimate of which share of a workweek a user typically spends using any given skill. Finally, we sum these skill vectors across all users in a city, providing us with an estimate of the mix of software skills a city hosts. Note that each number of these city-level skill vectors can be interpreted as the number of full-time equivalent software developers per skill.

To describe the skill profile of cities, it is often useful to assess to what extent a skill is overrepresented in a given city compared to its prevalence in the overall global economy. To do so, we once again rely on RCAs (equivalent to location quotients). In particular, city $c$'s RCA in skill $\theta$ is defined as:

\begin{equation}
    {RCA}_{c,\theta} = \frac{U_{c,\theta}/U_c}{U_{\theta}/U},
\end{equation}
where $U_{c,\theta}$ is the sum of skill shares in skill $\theta$ across all users that reside in city $c$ and omitted indices indicate summations over the corresponding dimension. That is, the RCA compares the share of full-time equivalent workers that city $c$ allocates to skill $\theta$ to the share that the world allocates to this skill. RCA values greater than one indicate that skill $\theta$ is overrepresented in the city, whereas values below one indicate that skill $\theta$ is underrepresented. In the Appendix we report the top five skills with the highest RCA in a selection of notable cities: San Jose (Silicon Valley), Berlin, Bengaluru, Beijing, Moscow, Vienna, and Rio de Janeiro.

\subsection{Skill relatedness}
Skills are often related to a specific subset of other skills, either because they are relatively similar, or because they often have to be combined to yield useful results. Although these are conceptually distinct types of relatedness, in both cases, we would expect that sets of related skills are disproportionately practiced by the same users \citep{anderson2017skill}. Exploiting this insight, we estimate the relatedness between two software skills by counting how often two skills are practiced by the same users. To do so, we first turn the elements of a user's skill vector into Pointwise Mutual Information (PMI) scores following \cite{van2023information}. 
PMI scores are similar to RCA values in that they help determine which skills a user disproportionately engages with on SO.\footnote{In fact, there is a close relation between RCA and PMI, where PMI is essentially the log-transformed RCA. Consequently, an RCA value of 1 corresponds to a PMI of 0, such that $PMI>0$ indicates overrepresentation of a skill in a user's SO activity. However, when estimating expected PMIs, we need to correct for the nonlinear way in which terms enter this expectation, which is why our estimates of PMI have additional terms from a Taylor expansion \citep[see][]{van2023information}} Next, we dichotomize these PMIs into skills that a user engages with (${PMI}>0$) and skills that a user does not significantly engage with (${PMI} \leq 0$). Finally, we construct a new network, this time of skills, counting the co-occurrences of skills in users, estimate expected PMIs and use these estimates as a measure of how related two skills are.

\subsection{The value of software skills}
To get a rough sense of how valuable different software skills are, we use the wage information provided by respondents to the annual Stack Overflow developer survey, as well as the technologies these respondents list. Many of these technologies are also found as tags on SO. \citet{feng2025building} leverage this  to connect users in SO probabilistically to one or more survey respondents. The strength of a link between an SO user and a survey respondent reflects the extent to which the tags in the post of the former coincide with the tags listed by the latter. Using such weights, we can calculate for each SO user the link-weighted average wage of the survey respondents they connect to.

To prevent differences in wage levels across countries from driving this estimate, we focus on US survey respondents only. These valuations should therefore not be interpreted as estimates of the actual wages of SO users  ---which will depend on labor market conditions and other factors--- but as valuations by the US software development labor market of their skills.\footnote{\citet{feng2025building} show that these skill value estimates also are meaningful outside the SO platform, acting as significant predictors of wage offers in a sample of online job ads for software developers.} We next calculate an estimated value of each skill as the weighted average of these imputed (user-level) valuations across all users, using the intensity with which a user engages with each skill as weights. Tables~\ref{tab:bottom_wages} and \ref{tab:top_wages} present the bottom and top~10 skills in terms of this value.

\singlespacing
\begin{table}
\centering
\begin{tabular}{rlr}
\toprule
Rank & Skill Description & Salary (\$) \\
\midrule
1 & Develop responsive WordPress themes with custom plugins and API integrations & \$98,408 \\
2 & Implement a dynamic PDF report generator for web content & \$105,580 \\
3 & Implement a responsive iframe embed for Vimeo videos & \$106,127 \\
4 & Develop a web application using Symfony framework and Doctrine ORM & \$106,191 \\
5 & Integrate various payment gateways in an e-commerce platform & \$107,652 \\
6 & Develop responsive UI with modern CSS frameworks and libraries & \$107,707 \\
7 & Implement a dynamic web form with various input elements & \$109,812 \\
8 & Implement a responsive sticky header with parallax scrolling effect & \$109,834 \\
9 & Develop dynamic web interfaces in JSF and Spring Webflow & \$109,873 \\
10 & Implement form submission with input validation & \$110,132 \\
\bottomrule
\end{tabular}
\caption{Skills with lowest inferred salaries.}\label{tab:bottom_wages}
\end{table}

\begin{table}
\centering
\begin{tabular}{rlr}
\toprule
Rank & Skill Description & Salary (\$) \\
\midrule
1 & Develop and optimize an iOS application user interface & \$152,063 \\
2 & Develop a mobile app with modern networking and UI frameworks & \$151,121 \\
3 & Develop a mobile app with offline data synchronization capabilities & \$150,377 \\
4 & Implement scalable real-time data processing pipelines & \$148,385 \\
5 & Implement server-side routing with Kotlin and ASP.NET & \$147,525 \\
6 & Develop a modern Android app with navigation and data management & \$147,376 \\
7 & Configure CI/CD pipeline with Kubernetes and Maven & \$147,067 \\
8 & Develop an Android app with modern UI/UX components & \$146,916 \\
9 & Implement parallel algorithms with performance metrics and visualization tools & \$146,179 \\
10 & Implement a location-based service application using MapKit & \$146,080 \\
\bottomrule
\end{tabular}
\caption{Skills with highest inferred salaries.}\label{tab:top_wages}
\end{table}
\onehalfspacing
\vspace{-5mm}
Low-value skills center on routine web development: WordPress theming, form handling, CSS layouts, iframe embedding, and e-commerce payment integration. These represent mature, widely diffused capabilities that have been commoditized over the past two decades. High-value skills, by contrast, involve mobile platform development with modern frameworks (iOS with SwiftUI, Android with Jetpack Compose), real-time data processing pipelines, CI/CD automation with containerization (Kubernetes), and parallel computing.

\section{Results}
\label{section4results}

\subsection{City rankings}

\begin{figure}
    \centering
    \includegraphics[width=\textwidth]{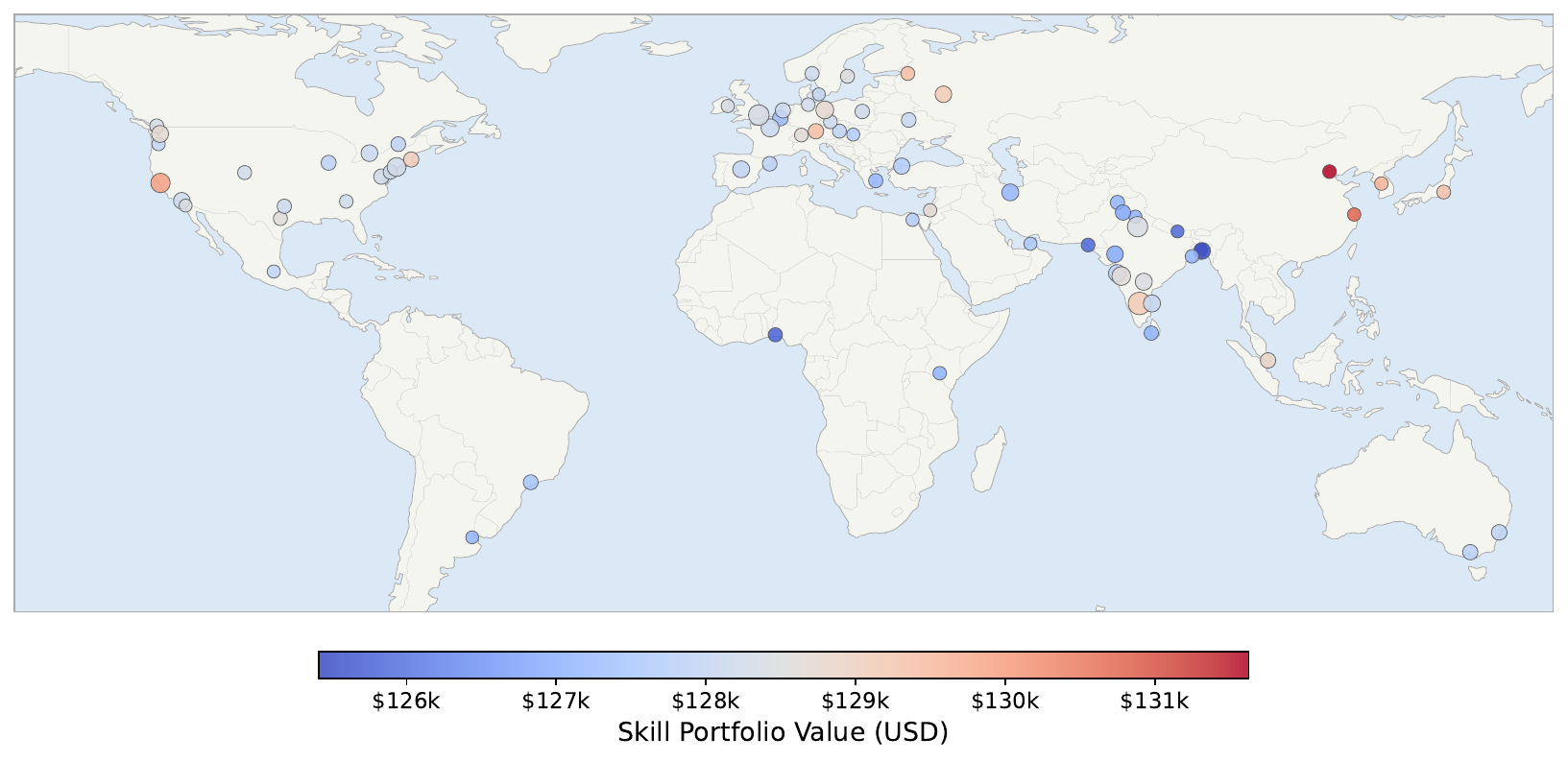}
    \caption{Average developer skill value across cities with at least 2,000 developers. Color indicates the weighted average salary implied by each city’s skill composition. Circle size is proportional to the number of developers.}
    \label{fig:city_values_map}
\end{figure}

Fig.~\ref{fig:city_values_map} maps the average skill value of cities with at least 2,000 full-time equivalent developers. We aggregate imputed skill values to the level of cities by summing skill vectors across all users, using posts and locations from 2019--2023. We use these summed skill shares as weights to calculate the weighted average skill value (based on salary data for 2022) in a city, providing us with an estimate of which cities engage in the most valuable software activities. Circle sizes are proportional to the number of developers. Tables~\ref{tab:cities_wages} and \ref{tab:bottom_cities_wages} in the Appendix report the top and bottom~20 cities among those with at least 500 FTE developers.

East Asian metropolises, led by Beijing, Shanghai, Seoul, and Tokyo, rank highest. Their developers concentrate in high-value mobile, systems, and machine learning skills, reflecting the orientation of major East Asian technology firms. North American and European hubs cluster in the middle range, while South Asian cities rank lower on average. San Jose ranks 7th, suggesting that the world’s leading software hub also carries out some lower-value activities alongside its frontier work.

The relatively compressed salary range (about 128,000 USD $\pm$ 3,000 USD) reflects the nature of our estimation strategy. Because we impute values from the skill composition of a city’s developer workforce, our measure captures differences in what kind of software work a city does, not differences in local wage levels. Actual developer salaries will vary substantially more across cities due to local labor market conditions, cost of living, and purchasing power differences. Our measure deliberately abstracts from these factors to isolate the compositional dimension of a city’s software portfolio.

The prominence of Chinese cities at the top of the ranking warrants interpretation. Shenzhen, Beijing, Hangzhou, and Shanghai all rank in the top ten, ahead of Silicon Valley. This partly reflects genuine specialization: Chinese tech firms have invested heavily in mobile development (both Android and iOS ecosystems), systems programming, and machine learning, all of which are high-value skills in our framework. At the same time, it likely also reflects a selection effect. Stack Overflow is an English-language platform, and Chinese developers have access to domestic alternatives such as CSDN and SegmentFault. Those who actively participate on SO may disproportionately work on international or frontier projects, inflating the apparent sophistication of Chinese cities’ skill portfolios. The ranking should therefore be read as describing the revealed skill composition of a city’s SO-active developers, which may not fully represent the local software sector as a whole.

\onehalfspacing
\subsection{Principle of relatedness}
The literature on the principle of relatedness \citep{hidalgo2018principle}  hypothesizes that cities enter into new activities that are closely related to the ones they are already active in. Does this also hold for software, a sector whose products are digital and in which collaboration is often organized on online platforms like GitHub? To answer this question, we follow the process outlined in \cite{li2024evaluating}: we first define a notion of relatedness and then calculate the density in terms of this relatedness around each skill in a city to quantify the amount of local activity in closely related skills. To be precise, we define the density $d_{c\theta}$ of a city $c$ around a given software skill, $\theta$:

$$d_{c\theta} = \sum_{\iota \neq \theta \in \Theta} \frac{\phi_{\theta \iota}}{\sum_{\kappa \neq \theta \in \Theta} \phi_{\theta \kappa}} \, \mathbf{1}\!\left[\mathrm{RCA}_{c\iota} \geq 1\right],$$
where $\Theta$ denotes the collection of all skills, $\phi_{\theta \iota}$ is a PMI-based measure of how related skill $\theta$ is to skill $\iota$, and $\mathbf{1}[\cdot]$ indicates that city $c$ is overspecialized (i.e., ${RCA}>1$) in skill $\iota$. The proximity weights are row-normalized, so $d_{c\theta} \in [0,1]$: a density of 0.4 means city $c$ is specialized in skills carrying 40\% of the relatedness mass around skill $\theta$. We measure density from city-skill profiles in the last year of the base period (2018).

\begin{figure}
    \centering
    \includegraphics[width=0.6\textwidth]{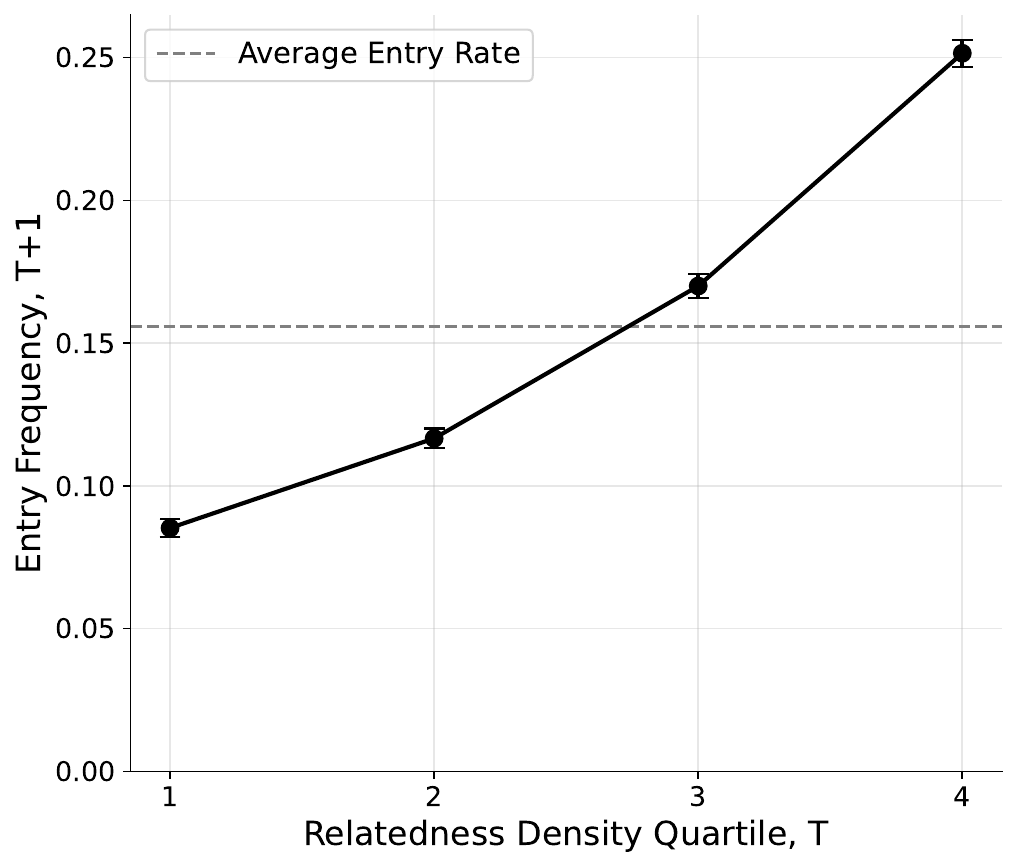}
    \caption{Relationship between relatedness density and entry in a new skill. The vertical axis plots the likelihood of a jump post-2018 in RCA from below 0.25 to above 1, the horizontal axis the city's relatedness density $d_{c\theta}\in[0,1]$ around the skill. Observations have been grouped by their density into four equally sized bins. Vertical lines denote 95\% confidence intervals.}\label{fig:entry}
\end{figure}

For each of the 237 skills, we construct a cross-section of all cities in our sample. These city-skill cross-sections are stacked to form the estimation sample, yielding approximately 214,000 observations (903 cities $\times$ 237 skills). To test whether density is predictive of future diversification, we study ``jumps'' in activity: events in which a city's RCA increases abruptly from below 0.25 to above 1. That is, we define diversification events as instances in which a city’s RCA in a skill was below 0.25 in the period 2008-2018 ($t_{0}$) and over 1 in the period 2019-2023 ($t_{1}$). We define $E$ (entry) as follows:

$$E_{c\theta} = 
\begin{cases} 
1 & \text{if } \text{RCA}_{c \theta t_{0}} \leq 0.25 \text{ and } \text{RCA}_{c \theta t_{1}} \geq 1 \\
0 & \text{if } \text{RCA}_{c \theta t_{0}} \leq 0.25 \text{ and } \text{RCA}_{c \theta t_{1}} < 1\\
\text{missing} & \text{otherwise} 
\end{cases}$$
To assess the density in the city around the skill, we use city-skill profiles from the last year of the base period, $t_0$ (i.e., from 2018). We limit our sample to city-skill combinations that are at risk of being entered, i.e., for which the base-period RCA is below 0.25. Next, we calculate for these cells the density of skills that surround them. We then estimate for each quartile of city $c$'s density around a skill $\theta$ the likelihood of observing an entry into skill $\theta$. That is, we calculate the relative entry frequencies within each density quartile. This is plotted in Fig. \ref{fig:entry}.

Fig.~\ref{fig:entry} strongly supports our hypothesis that cities diversify their software activities by moving into related skills. Moving from the lowest to the highest bin increases the likelihood of observing an entry by nearly 50\%. In Table \ref{tab:reg_entry}, we analyze this question more rigorously by estimating linear probability models of the form:
\begin{equation}\label{eq:entry_lpm}
E_{c\theta} = \beta\, d_{c\theta} + \mathbf{X}_{c\theta}'\boldsymbol{\gamma} + \alpha_c + \delta_\theta + \varepsilon_{c\theta},
\end{equation}
where $E_{c\theta}$ is the entry indicator defined above, $d_{c\theta}$ is the relatedness density (levels), $\mathbf{X}_{c\theta}$ is a vector of controls (base-period user counts, a zero indicator), $\alpha_c$ and $\delta_\theta$ are city and skill fixed effects (included in later columns), and $\varepsilon_{c\theta}$ is the error term. Controls include the size of the skill in the city in the base period, a dummy for whether the skill was wholly absent from the city in the base year, and the total number of SO users in the city and in the skill globally. We cluster standard errors by city and skill.

The results corroborate the findings in Fig.~\ref{fig:entry}. Our preferred specification (model 6) suggests that a one-standard-deviation increase in relatedness density (0.10 on the [0,1] scale) raises the likelihood of entry by about 5.0 percentage points, against an average entry rate of 15.6\%. Moving from the 10th to the 90th percentile of density (roughly 0.03 to 0.27) raises the entry probability by about 12 percentage points. %The density coefficient is stable when we replace the (log) FUA population control with the (log) number of SO developers in the city, which captures the size of the local software sector rather than the city as a whole (Tables~\ref{tab:entry_lpm_so_devs} and~\ref{tab:ppml_so_devs} in the Appendix).

\rowcolors{1}{white}{white} %
\begin{table}
\centering
\begin{tabular}{lcccccc}
   \toprule
   & \multicolumn{6}{c}{Entry: RCA $\leq 0.25$ to $\geq 1$}\\
                                       & (1) & (2) & (3) & (4) & (5) & (6)\\
   \midrule
   Relatedness Density at T  & 0.683$^{***}$ & 0.684$^{***}$ & 0.707$^{***}$ & 0.636$^{***}$ & 0.684$^{***}$ & 0.511$^{***}$\\
                                       & (0.020) & (0.021) & (0.020) & (0.040) & (0.020) & (0.032)\\
   (Log) N users city C, skill Y, time T & & -0.069$^{***}$ & -0.085$^{***}$ & -0.082$^{***}$ & -0.078$^{***}$ & -0.077$^{***}$\\
                                       & & (0.009) & (0.009) & (0.011) & (0.009) & (0.011)\\
   Zero indicator for log(0) & & 0.121$^{***}$ & 0.202$^{***}$ & 0.208$^{***}$ & 0.186$^{***}$ & 0.203$^{***}$\\
                                       & & (0.019) & (0.019) & (0.023) & (0.019) & (0.023)\\
   (Log) FUA Population & & & 0.008$^{***}$ & & 0.009$^{***}$ & \\
                                       & & & (0.002) & & (0.002) & \\
   (Log) N Users, skill Y, time T & & & 0.039$^{***}$ & 0.041$^{***}$ & & \\
                                       & & & (0.003) & (0.003) & & \\
   \midrule
   Observations & 131,257 & 131,257 & 131,257 & 131,255 & 131,257 & 131,255\\
   R$^2$ & 0.034 & 0.034 & 0.048 & 0.06 & 0.071 & 0.083\\
   \midrule
   City FE & & & & Y & & Y\\
   Skill FE & & & & & Y & Y\\
   \bottomrule
\end{tabular}

\caption{Linear probability models predicting entry of a city into a software skill, defined as having an RCA below 0.25 in 2008--2018 to above 1 in 2019--2023. FUA population is taken from \cite{schiavina2019ghsfua}. Standard errors are clustered on city and skill.}\label{tab:reg_entry}
\end{table}

We also analyze how the number of users with a particular skill in a city rises with density, using Poisson Pseudo Maximum Likelihood (PPML) estimation. Results (Table~\ref{reg:PPML} in the Appendix) corroborate the importance of density: in our preferred specification a one-standard-deviation increase in relatedness density raises the expected number of software developers in a city-skill pair by about 20\%. Altogether the descriptive figure and the regression analyses present strong support for our first hypothesis: the spatial dynamics of the software sector are substantially constrained by the principle of relatedness, such that cities move preferentially into related skills. This also shows that relatedness operates at the fine-grained level of skills beyond the aggregate levels of sectors, industries, and occupations. 

\subsection{Diffusion of new skills}\label{sec:newskills}
The level of detail of our skill data allows us to identify newly emerging skills. We define a skill as emerging if less than 15\% of its total posting content appears before 2014. Out of a total of 237, this yields 35 emerging skills. Examples include skills related to developer operations (DevOps) and testing, deep learning and AI, big data engineering, cloud computing, and blockchain. A full list is provided in the Appendix, Table~\ref{tab:emerging_skills}.

For each emerging skill, we designate a single city of origin. We first determine a breakthrough year: the first year in which cumulative global activity in the skill exceeds 5\% of its eventual peak. We then identify which cities were active in the skill during this early period. To identify origins, we iterate year by year from the earliest data. In each year, we check whether any city meets a double criterion: RCA~$\geq$~1 \emph{and} at least three full-time equivalent developers in the skill. The first year in which any city qualifies determines the origin cohort. Among cities qualifying in that first year, we break ties by highest RCA, favoring early specialization over sheer size. If no city meets the threshold in any single year, we fall back to the city with the highest cumulative activity. This first-to-qualify procedure avoids designating large cities as origins simply because they accumulate more total activity over time. For each of the 35 emerging skills, we construct a cross-section in which every city other than the origin is a potential adopter (approximately 975 cities per skill). These cross-sections are stacked to form the estimation sample.

\begin{figure}
\begin{center}
    \includegraphics[width=0.9\linewidth]{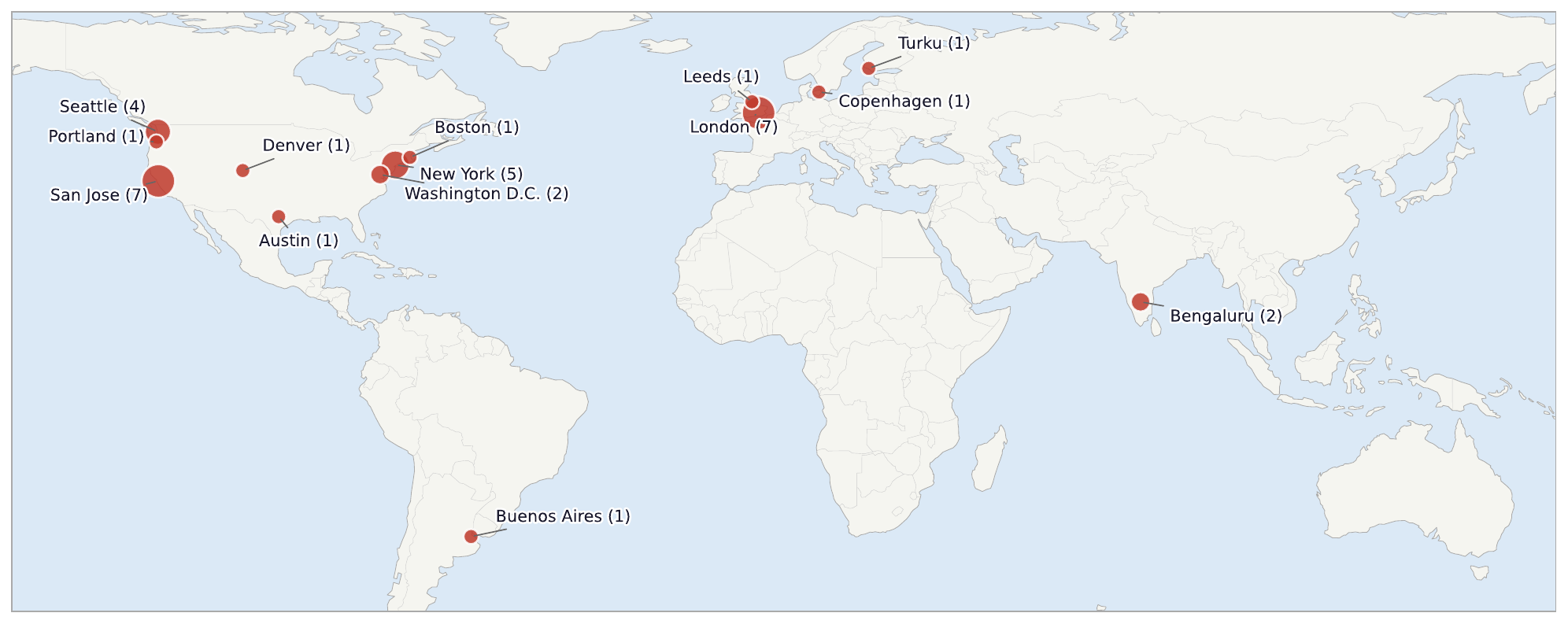}
\end{center}
    \caption{Origin cities for emerging software skills. Circle sizes are proportional to the number of skills originating in each city.}
    \label{fig:InnovatorsMap}
\end{figure}

Table~\ref{tab:new_task_cities_expanded} counts how often each city serves as the origin of an emerging skill. London and Silicon Valley (San Jose) are jointly the most important origins, each accounting for 7 of 35 emerging skills. New York follows with 5 and Seattle with 4. Bengaluru and Washington D.C. each originate 2 skills, and eight further cities originate one apiece. The 35 origins are distributed across just 14 distinct cities, with four of them (London, San Jose, New York, Seattle) accounting for nearly two-thirds of all skill origins. Fig.~\ref{fig:InnovatorsMap} maps these origin cities, with circle sizes proportional to the number of skills originating there.

\singlespacing
\begin{table}[b]
\centering
\rowcolors{2}{gray!25}{white}
\begin{tabular}{ll ll}
        \toprule
        City & Count & City & Count \\
        \midrule
London & 7 & Austin & 1 \\
San Jose & 7 & Portland & 1 \\
New York & 5 & Buenos Aires & 1 \\
Seattle & 4 & Copenhagen & 1 \\
Bengaluru & 2 & Leeds & 1 \\
Washington D.C. & 2 & Denver & 1 \\
Boston & 1 & Turku & 1 \\
\bottomrule
    \end{tabular}
\caption{Origin cities for emerging software skills. Each emerging skill is assigned a single origin: the city with the highest early-period activity that meets a double criterion (RCA~$\geq$~1 and $\geq$~3 FTE developers).}
\label{tab:new_task_cities_expanded}
\end{table}
\onehalfspacing

\vspace{-5mm}
Next, we ask what predicts which cities become origin locations for new skills. The question is a city-level one: what kinds of cities generate new skills in general. We therefore take the city as the unit of observation and model the count of emerging skills a city originates:
\begin{equation}\label{eq:emergence}
n_c = f\!\left(\beta_1 \log P_{c} + \beta_2 H_{c}\right),
\end{equation}
where $n_c$ is the number of emerging skills for which city $c$ is the designated origin, $P_c$ is FUA population taken from \cite{schiavina2019ghsfua}, and $H_c$ is the Shannon entropy of the city's base-period (2008--2018) skill distribution, which captures diversity within the city's software sector. We estimate three specifications: an OLS regression of the count $n_c$, a linear probability model of an indicator for whether the city originates any emerging skill, and an OLS regression of $\log(1+n_c)$. The sample consists of the 976 cities for which we observe both SO activity and population.

Table~\ref{tab:emergence} reports the results. Greater population size and skill diversity are both associated with a higher chance that a city originates new skills. In the linear probability specification (column~4), moving from the 10th to the 90th percentile in a city's population size raises the probability that a city brings forth at least one new skill by about 3.7 percentage points, and a corresponding move in skill diversity by about 3.3 percentage points. Against a baseline rate of 1.4\% of cities that act as origins for any new skills, these are sizeable effects. The OLS count specifications (columns~1--3) and the $\log(1+n_c)$ specification (column~5) concur, with significant positive coefficients on population size and skill     diversity.

These findings confirm hypothesis~\ref{hyp:emergence}: new skills emerge in large cities with diversified software sectors, consistent with the nursery-cities view. Because the 35 origins are concentrated in 14 distinct cities, with London and San Jose each accounting for seven, we re-estimate all five specifications dropping these top contributors separately and jointly (Table~\ref{tab:emergence_robustness} in the Appendix). Both coefficients remain positive and significant across all three subsamples.

\rowcolors{1}{white}{white}
\begin{table}
\centering
\renewcommand\cellalign{t}
\begin{threeparttable}
\begin{tabular}{lccccc}
\toprule
 & \multicolumn{3}{c}{No.\ skills originated (OLS)} & \makecell{Originates any\\skill (LPM)} & \makecell{$\log(1+$No.$)$\\(OLS)} \\
\cmidrule(lr){2-4} \cmidrule(lr){5-5} \cmidrule(lr){6-6}
 & (1) & (2) & (3) & (4) & (5) \\
\midrule
Log FUA population & \makecell{0.049$^{*}$ \\ (0.020)} &  & \makecell{0.042$^{*}$ \\ (0.018)} & \makecell{0.012$^{**}$ \\ (0.004)} & \makecell{0.016$^{**}$ \\ (0.006)} \\
\addlinespace
Skill diversity (entropy) &  & \makecell{0.132$^{**}$ \\ (0.047)} & \makecell{0.086$^{**}$ \\ (0.029)} & \makecell{0.038$^{***}$ \\ (0.010)} & \makecell{0.039$^{***}$ \\ (0.011)} \\
\addlinespace
Intercept & \makecell{-0.628$^{*}$ \\ (0.260)} & \makecell{-0.568$^{**}$ \\ (0.205)} & \makecell{-0.926$^{**}$ \\ (0.353)} & \makecell{-0.323$^{***}$ \\ (0.091)} & \makecell{-0.385$^{**}$ \\ (0.124)} \\
\midrule
Observations & 976 & 976 & 976 & 976 & 976 \\
$R^2$ & 0.022 & 0.012 & 0.027 & 0.034 & 0.034 \\
\bottomrule
\end{tabular}

\footnotesize Significance: $*$ p $<$ 0.05, $**$ p $<$ 0.01, $***$ p $<$ 0.001. Robust standard errors in parentheses.
\end{threeparttable}
\caption{ City-level emergence regressions. The dependent variable is the count of emerging skills a city originates (columns~1--3), an indicator for originating any emerging skill (column~4), or $\log(1+\text{count})$ (column~5).}
\label{tab:emergence}
\end{table}

\subsection{Windows of locational opportunity}\label{sec:wlo}

Even though most new skills emerge in well-established software hubs, their location patterns may still be less spatially constrained than those of existing skills. WLO theory holds that at an industry's inception it is often unclear what production factors are required, and the geographical conditions a new activity needs to thrive may not yet exist anywhere \citep{boschma1997new,walker1989capitalist}. We do not study the emergence of whole new industries but of new skill sets within an existing software sector, so we do not expect overwhelming WLO effects. Yet new skills may still emerge more freely than existing skills diffuse.

\begin{figure}
\centering
\includegraphics[width=\linewidth]{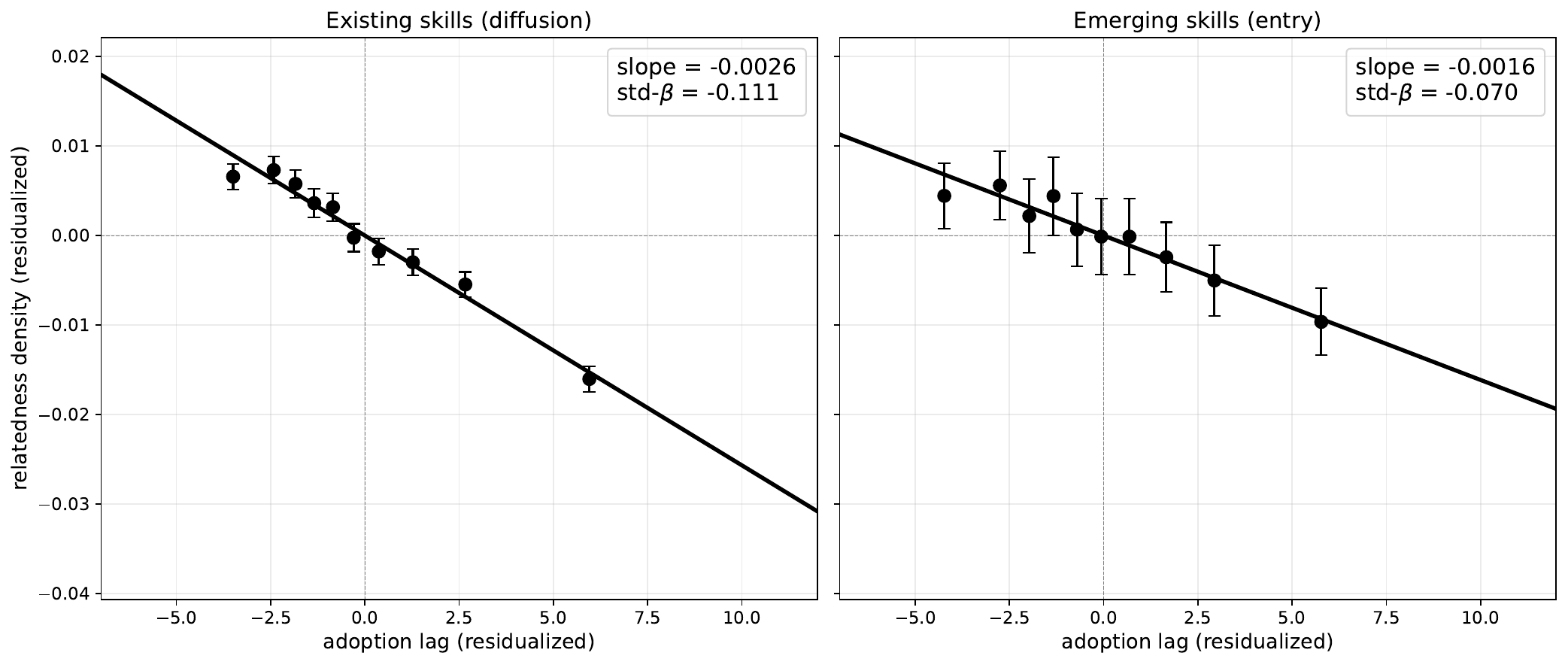}
\caption{Relatedness density and adoption lag, residualized on city and skill fixed effects; points are binned city-skill scatters with 95\% confidence intervals. Left: existing skills, Right: emerging skills. The shallower emerging slope is the direction predicted by WLO. A pooled interaction test of differences between the slopes gives $p \approx 0.055$.}\label{fig:wlo}
\end{figure}

We test this by comparing the relatedness-density gradient on emerging-skill entry against that on existing-skill diffusion (Fig.~\ref{fig:wlo}). For each city--skill pair we take one cross-section at a baseline year and measure how the city's  relatedness density at that baseline predicts the lag until the city reaches RCA $\geq 1$ in the skill. For emerging skills the baseline is the skill's global breakthrough year (median 2011); for existing skills we use 2011 as a common matched baseline. Both panels subtract city and skill fixed effects.

The gradient is shallower for emerging than for existing skills, in the direction WLO predicts. In standardized terms the emerging slope is about 37\% less steep than the existing slope (std-$\beta$: $-0.070$ vs.\ $-0.111$). A pooled interaction test under the same clustering yields $\beta_{\mathrm{lag}\times\mathrm{emerging}} = +0.0010$ (SE $0.0005$, $p \approx 0.055$; Table~\ref{tab:wlo_interaction} in the Appendix). The small number of emerging skills ($n=35$) limits power for this formal test.

We read this as support for hypothesis~\ref{hyp:WLO}: local relatedness governs the diffusion of existing skills somewhat more tightly than the emergence of new ones, though even brand-new skills emerge preferentially in cities that already developed many related skills.

\subsection{Diffusion of emerging skills}
We now turn to how emerging software skills diffuse beyond their places of origin. In this analysis, every city other than the origin is a potential adopter. To assess the importance of geography on the diffusion of new skills, we estimate Cox proportional hazard models of the form:
\begin{equation}\label{eq:cox}
h(t \mid \mathbf{z}_{c\theta}) = h_0(t) \exp\!\left(\beta_1 \log D_{c\theta} + \beta_2\, d_{c\theta} + \mathbf{X}_{c}'\boldsymbol{\gamma}\right),
\end{equation}
where $h(t \mid \mathbf{z}_{c\theta})$ is the hazard of city $c$ adopting skill $\theta$ at time $t$, $h_0(t)$ is the baseline hazard, $D_{c\theta}$ is the geodesic distance between city $c$ and the skill's origin, $d_{c\theta}$ is the relatedness density around the new skill, and $\mathbf{X}_c$ includes city size and software skill diversity. Relatedness density is computed from city-skill profiles at the skill's breakthrough year, so that the measure reflects the capability landscape at the time of emergence rather than a pooled average across the training period. The dependent variable is the time from a skill's breakthrough year to the first year in which a city achieves RCA~$>$~1 in that skill. Cities that never adopt by the end of our observation window (2023) are right-censored. This survival framework thus exploits the full temporal variation in adoption timing, rather than reducing adoption to a binary indicator.

The setup allows us to pit two fundamentally different forces against each other. The first is supra-local: geographic distance between a city and the skill's origin, which captures the classic diffusion channel emphasized by ILC theory. If tacit knowledge matters, distance from the origin should slow adoption. The second force is hyper-local: relatedness density, computed entirely from a city's own capability profile. If the principle of relatedness applies, cities with stronger existing capabilities in related skills should adopt faster, regardless of where those skills originated. Distance enters in logs, while relatedness density enters as a level on the interval $[0,1]$. The distance coefficient is therefore a hazard semi-elasticity, and the density coefficient is the proportional change in the hazard per unit (0 to 1) of density. Results are shown in Table~\ref{tab:survival}. As a robustness check, we also estimate linear probability models (Table~\ref{tab:horserace} in the Appendix) that model adoption likelihoods within a 3-year window after the skill first emerged. These models yield qualitatively identical conclusions.

\paragraph{Relatedness density dominates distance.} Relatedness density is by far the dominant predictor of adoption speed. The coefficient on relatedness density is $4.85$ (column~3): a one-standard-deviation increase ($\approx 0.11$ on the $[0,1]$ scale) raises the adoption hazard by about 74\%, and by about 34\% in the full specification with population and entropy controls (column~5). Distance to the origin city is either quantitatively small or not significant, implying that distance to the origin has no meaningful effect on adoption speed. These results are robust to skill stratification (Column~6), which allows baseline adoption rates to vary freely across skills.\footnote{For context, gravity models in international trade typically estimate distance elasticities between $-0.7$ and $-1.5$ \citep{head2014gravity}; our estimate is more than an order of magnitude smaller.}

\paragraph{Concrete distance effects.} To illustrate the magnitude, consider a new skill originating in San Jose. The distance to Seattle is approximately 1,200~km, to New York 4,100~km, to London 8,600~km, and to Bengaluru 13,800~km. Across specifications that control for relatedness density, the coefficient for log-distance ranges from $-0.006$ to $-0.055$. Even at the upper bound, moving from Seattle-distance to Bengaluru-distance from San Jose changes the adoption hazard by only $13\%$. By comparison, a one-standard-deviation increase in relatedness density raises the hazard by roughly 34--74\% depending on the specification. Whether a new skill reaches Bengaluru quickly depends on whether Bengaluru already has developers working on related skills, not on the 14,000~km separating it from Silicon Valley.

\paragraph{City characteristics.} Column~5 adds city size and skill diversity as controls. Larger cities adopt faster ($0.080$, $p < 0.001$) consistent with thicker labor markets facilitating absorption of new capabilities. Cities with more diverse skill portfolios also adopt faster ($1.180$, $p < 0.001$). Even after controlling for city size and diversity, the distance effect remains significant but substantively small ($-0.031$, $p < 0.001$), confirming that local capabilities matter far more than proximity to the origin. %Population and the SO developer count capture overlapping dimensions of the local economy, and substituting one for the other does not change the qualitative conclusion. 
These findings strongly corroborate hypothesis~\ref{hyp:horserace}: in software, a city's ability to adopt new skills depends overwhelmingly on its related capabilities, not on its proximity to where the skill first emerged.

\begin{table}
\begin{threeparttable}
\begin{tabular}{lcccccc}
\toprule
 & \multicolumn{6}{c}{adoption hazard} \\
\cmidrule(lr){2-7}
 & (1) & (2) & (3) & (4) & (5) & (6) \\
\midrule
Log distance to origin & \makecell{-0.068*** \\ (0.009)} &  & \makecell{-0.024*** \\ (0.007)} & \makecell{-0.055*** \\ (0.007)} & \makecell{-0.031*** \\ (0.006)} & \makecell{-0.006 \\ (0.008)} \\
\addlinespace
Relatedness density &  & \makecell{4.873*** \\ (0.084)} & \makecell{4.851*** \\ (0.083)} & \makecell{4.413*** \\ (0.091)} & \makecell{2.573*** \\ (0.091)} & \makecell{4.924*** \\ (0.086)} \\
\addlinespace
Log population &  &  &  & \makecell{0.114*** \\ (0.010)} & \makecell{0.080*** \\ (0.008)} &  \\
\addlinespace
Skill diversity (Shannon entropy) &  &  &  &  & \makecell{1.180*** \\ (0.036)} &  \\
\addlinespace
\midrule
Skill strata & - & - & - & - & - & x \\
\midrule
Observations & 34027 & 34027 & 34027 & 34027 & 34027 & 34027 \\
Events & 25853 & 25853 & 25853 & 25853 & 25853 & 25853 \\
Concordance & 0.532 & 0.693 & 0.693 & 0.695 & 0.708 & 0.692 \\
\bottomrule
\end{tabular}

\caption{Cox proportional hazard models predicting time to adoption of emerging skills. One origin per skill; all other cities are potential adopters. Reported coefficients are log-hazard ratios. Positive coefficients indicate faster adoption (higher hazard). Standard errors clustered by city.}
\label{tab:survival}

\end{threeparttable}
\end{table}

\section{Discussion}\label{section5discussion}

Today the spatial distribution of human capital, expertise and skills, is a decisive factor in the divergence of local economies worldwide. Yet the newest and most quickly evolving skills are typically hardest to observe. As a consequence, we often have much better estimates of how economies diversify within a set of traditional economic activities than how economies diversify into activities that are still emerging. It is precisely these emerging activities that policymakers, but also firms and workers have a keen interest in. Currently, one of the most rapid transformations of work can be found in the area of software.%Over the course of less than two decades, the number of users in our dataset grew from around 20 thousand to over 6 million. 

In this paper, we have studied 3M geolocated software developers on Stack Overflow in terms of the types of software skills they engage with, through the lens of a taxonomy of software skills analogous to existing datasets of professional skills such as O*NET in the US or ESCO in Europe. We use this taxonomy to analyze the spatial distribution of software skills, how cities diversify within this space, where new software skills emerge and how they diffuse beyond these initial locations. 

Software is often considered to be a special type of traded activity, whose products are ``weightless’’. In this sense, it is unclear to what extent traditional economic geography frameworks apply to this sector. Here, we analyzed two workhorse frameworks from evolutionary economic geography: the principle of relatedness and the product or industry life cycle framework. Whereas the former is ``hyper-local,'' explaining a city's transformation trajectory purely based on what happens within a city, the latter also describes the opposite, diffusion between cities. We find that both frameworks prove relevant to the spatial dynamics of the software industry. However, the evidence is much stronger for hyper-local, within-city forces. In line with the principle of relatedness, we find that cities preferentially diversify into new software skills that are closely related to skills in which their software developers already had ample experience. Moreover, in line with ILC and nursery cities arguments, new skills tend to emerge in large cities with diversified software sectors. Both components of the classical nursery-cities prediction hold: city size and diversity each predict the generation of new skills, regardless of whether these variables enter the analysis individually or jointly. In contrast, we find only very mild effects of distance in constraining the diffusion of these new skills. Survival models estimating the speed of adoption confirm this finding: relatedness density strongly predicts adoption speed while geographic distance has at most a small effect. In line with WLO arguments, the \emph{relatedness-density} constraint on a city's entry into a new skill is weaker when the skill is new than when it is established. 

These results point to a two-stage process in the spatial evolution of the software industry. In the first stage, new skills are generated in large and diversified software ecosystems, consistent with the classical nursery-cities mechanism \citep{jane1969economy,duranton2001nursery}: novelty thrives where many developers work across a wide range of specializations and where the resulting ``recombinant search'' across unusual skill combinations has a higher chance of producing a breakthrough. Here windows of opportunity are open for a larger set of regions: within this nursery phase there is room for non-incumbent cities to participate. In the second stage, once a skill's production recipe has been validated and codified, its diffusion follows the principle of relatedness: cities adopt new skills when they already possess closely related capabilities. This asymmetry between generation and adoption echoes the distinction in evolutionary theory between variation and selection. Generation benefits from the combination of scale and diversity that large urban software ecosystems provide. Adoption, by contrast, operates on fitness in a local capability niche, consistent with Marshallian specialization advantages \citep{duranton2001nursery}.

%At the same time, however, the geography of software activities looks surprisingly familiar to scholars that study the transformation of urban economies. Cities’ diversification in software activities follows the same path dependencies that have been observed for more traditional economic sectors. That is, cities’ software sectors evolve through related diversification, adding new tasks that are close in skill space to the activities they are already active in, further corroborating the findings on the principle of relatedness \citep{hidalgo2018principle}. 

Our work has a number of limitations. First, SO users represent just the tip of a much larger iceberg of software developers and SO usage may imperfectly capture the full population of programmers in the world economy. \cite{feng2025building} show that the skills identified on SO are also a reflection of software skills in the wider economy. Moreover, they find that the geographical distribution of SO users closely mimics that of GitHub contributors. However, the geographical usage patterns of the software skills we analyze may still be imperfectly captured by the spatial distribution of SO users. It is \emph{a priori} unclear how any biases in SO coverage would affect the hypothesis tests presented above but this would be an important area for future research. Moreover, Stack Overflow usage has recently plummeted with the arrival of generative AI based coding assistants \citep{del2024large}. We therefore plan to expand our analysis in future work using other types of data, such as job ads and coding activity on GitHub. 

Second, with the data at hand, we could only arrive at indicative estimates of the value of each software development skill. Additional data from job ads and their wage offers would also help us better distinguish between high and low value skills. Nevertheless, our survey based skill valuations already allowed us to identify a global division of labor in software in which lower-value, more standardized software skills are often produced in lower wage cities in developing countries, whereas high value software skills concentrate in some of the most expensive cities on earth. This may add an interesting dimension to the work on Global Production Networks \citep[e.g.,][]{yeung2015toward}.

Third, in a sector as dynamic as the software industry, we would expect skills to change. In future work, we therefore aim to analyze changes in software skills themselves. This would also allow us to analyze where not just new skills but also new bundles of skills---new jobs---emerge. 

Fourth, the focus on individuals leaves unexplored dimensions of teamwork and collaborative software development \citep{betti2025dynamics}, either in open source projects or within the institutional framework set by firms \citep{petralia2025open}. In future work, we aim to analyze such collaborations as well as their importance in firm performance by analyzing additional digital traces of software development, such as those recorded on GitHub.  

Finally, the methodology developed in this paper need not be limited to the world of software. Online question-answer platforms have arisen in several areas, from mathematics and physics research to cooking and gardening, and can be turned into skill datasets according to the same logic developed in this paper. 

% policy
\subsection{Policy and societal implications}
Our work also provides potentially useful information for policymakers. The principle of relatedness has started gaining prominence in policy debates at the local, national and supranational levels. The fact that this principle also predicts diversification in local software clusters suggests that the emerging policy frameworks based on this principle \citep{balland2019smart,boschma2015relatedness,li2024evaluating} can be extended to software development activity. Consequently, the highly granular analysis of software development work that we unlock could be used to help countries and cities around the globe not only map and benchmark their software capabilities against one another, but also chart feasible diversification paths into new software development activities.

% future research

%Furthermore, developers and firms moving to new cities likely influence and are influenced by the skills that are present there \citep{neffke2018agents,elekes2019foreign}, and transmit some knowledge back to their origins \citep{agrawal2006gone}. Sudden changes causing changes in mobility of developers, for instance following the Russian invasion of Ukraine \citep{wachs2023digital}, or due to changes in immigrant regulations \citep{khanna2017boom,brinatti2023third}, can be exploited to study the consequences of mobility more precisely.

Similarly, our analysis of skill relatedness is relevant for programmers who want to widen their skill profile and for firms that want to map career trajectories of their workers or help create teams that take such career paths into consideration. A better understanding of the geography of these skills  may inform educational tracks in local universities or vocational training programs. On the other hand, it may help support programmers who plan or need to make career changes or relocate to places where their skills are in demand.

Finally, our work offers a new perspective on the future of work, a field that currently is widely debated in academic and policy circles. The future of work is typically studied through the lens of computers and AI replacing or enhancing regular skills of workers. However, one particular way in which the nature of work changes is that more and more jobs require some level of coding skills. Just as computer skills became widespread in office work in the 1980s and 1990s, some level of programming skills may diffuse across an increasing range of jobs. Examples are already visible in instances such as data journalism, where large newspapers expand their teams with workers who are skilled at data analysis and visualization. Other information-facing jobs, such as polling and marketing, may start requiring more and more coding skills. 

In this sense, the developments in software are not only impacting on the future of work, software skills themselves may be part of this future. At the same time, AI coding assistants complicate this picture: they make basic programming more accessible, which may broaden the set of jobs that touch code, while also raising the premium on developers who can supervise and integrate AI-generated code in production systems. A better understanding and measurement of these phenomena is indispensable for workers, firms and policymakers that need to adapt to a rapidly changing nature of work.

\subsection*{Acknowledgements}
JW acknowledges support from the Hungarian National Scientific Fund (EXCELLENCE 149614 and OTKA FK 145960). XF, SD, and FN acknowledge support from the Austrian Research Promotion Agency (FFG): 873927 (ESSENCSE).

\bibliography{references}

\clearpage % Start the table on a new page
\section*{Appendix}
\subsection*{Robustness: Linear probability models for diffusion}

As a robustness check on the survival analysis in Table~\ref{tab:survival}, we estimate linear probability models predicting a binary adoption indicator (whether a city achieves RCA~$>$~1 within three years of breakthrough). Results in Table~\ref{tab:horserace} are qualitatively similar to the Cox Proportional Hazard estimates. Relatedness density dominates distance by roughly 4:1 in the full specification with city and skill fixed effects.

\begin{table}[h!]
\centering
\renewcommand\cellalign{t}
\resizebox{\linewidth}{!}{
\begin{tabular}{lcccccc}
\toprule
 & \multicolumn{6}{c}{City is early adopter} \\
\cmidrule(lr){2-7}
 & (1) & (2) & (3) & (4) & (5) & (6) \\
\midrule
Log distance to origin (z) & \makecell{$-$0.026$^{**}$ \\ (0.010)} &  & \makecell{$-$0.008 \\ (0.007)} & \makecell{$-$0.003 \\ (0.005)} & \makecell{$-$0.014 \\ (0.010)} & \makecell{$-$0.008$^{*}$ \\ (0.004)} \\
\addlinespace
Relatedness density (z) &  & \makecell{0.200$^{***}$ \\ (0.007)} & \makecell{0.199$^{***}$ \\ (0.007)} & \makecell{0.197$^{***}$ \\ (0.008)} & \makecell{0.126$^{***}$ \\ (0.027)} & \makecell{0.072$^{***}$ \\ (0.013)} \\
\midrule
Skill FE & $-$ & $-$ & $-$ & x & $-$ & x \\
Target city FE & $-$ & $-$ & $-$ & $-$ & x & x \\
\midrule
Observations & 32,056 & 32,056 & 32,056 & 32,056 & 32,056 & 32,056 \\
$R^2$ & 0.003 & 0.185 & 0.185 & 0.236 & 0.229 & 0.287 \\
\bottomrule
\end{tabular}
}
\caption{Linear probability models predicting diffusion of emerging skills. Single-origin design: one origin per skill, all other cities are potential adopters. All variables standardized. Standard errors clustered by target city and skill.}\label{tab:horserace}
\end{table}

\clearpage
\subsection*{Robustness: Excluding San Jose as origin}

San Jose accounts for 7 of 35 skill origins, raising the question of whether our diffusion results are driven by a single dominant hub. Tables~\ref{tab:survival_no_sj} and \ref{tab:horserace_no_sj} re-estimate the survival and LPM models after dropping all observations where San Jose is the origin city. The sample drops from roughly 34,000 to 27,000 observations. The core finding is robust: relatedness density remains the dominant predictor of adoption speed. The distance coefficient is somewhat larger in magnitude in this subsample ($-0.032$ in column~3 vs.\ $-0.024$ in the full sample). Concordance indices quantifying the goodness of fit of these analyses are comparable or slightly higher than in the full sample.

\begin{table}[h!]
\begin{tabular}{lcccccc}
\toprule
 & \multicolumn{6}{c}{adoption hazard} \\
\cmidrule(lr){2-7}
 & (1) & (2) & (3) & (4) & (5) & (6) \\
\midrule
Log distance to origin & \makecell{-0.073*** \\ (0.009)} &  & \makecell{-0.032*** \\ (0.007)} & \makecell{-0.062*** \\ (0.007)} & \makecell{-0.039*** \\ (0.007)} & \makecell{-0.016 \\ (0.008)} \\
\addlinespace
Relatedness density &  & \makecell{4.745*** \\ (0.086)} & \makecell{4.715*** \\ (0.086)} & \makecell{4.324*** \\ (0.093)} & \makecell{2.547*** \\ (0.097)} & \makecell{4.758*** \\ (0.097)} \\
\addlinespace
Log population &  &  &  & \makecell{0.105*** \\ (0.010)} & \makecell{0.068*** \\ (0.008)} &  \\
\addlinespace
Skill diversity (Shannon entropy) &  &  &  &  & \makecell{1.165*** \\ (0.036)} &  \\
\addlinespace
\midrule
Skill strata & - & - & - & - & - & x \\
\midrule
Observations & 27216 & 27216 & 27216 & 27216 & 27216 & 27216 \\
Events & 20786 & 20786 & 20786 & 20786 & 20786 & 20786 \\
Concordance & 0.535 & 0.693 & 0.693 & 0.693 & 0.706 & 0.689 \\
\bottomrule
\end{tabular}
\caption{ Cox proportional hazards. Positive coefficients = faster adoption.
 Cluster-robust SEs (city). San Jose excluded as origin.
 Significance: $*$ p $<$ 0.05, $**$ p $<$ 0.01, $***$ p $<$ 0.001.}
 \label{tab:survival_no_sj}
\end{table}

\begin{table}[h!]
\centering
\renewcommand\cellalign{t}
\resizebox{\linewidth}{!}{
\begin{tabular}{lcccccc}
\toprule
 & \multicolumn{6}{c}{City is early adopter} \\
\cmidrule(lr){2-7}
 & (1) & (2) & (3) & (4) & (5) & (6) \\
\midrule
Log distance to origin (z) & \makecell{$-$0.026$^{*}$ \\ (0.011)} &  & \makecell{$-$0.009 \\ (0.008)} & \makecell{$-$0.005 \\ (0.006)} & \makecell{$-$0.014 \\ (0.011)} & \makecell{$-$0.009 \\ (0.005)} \\
\addlinespace
Relatedness density (z) &  & \makecell{0.198$^{***}$ \\ (0.008)} & \makecell{0.197$^{***}$ \\ (0.009)} & \makecell{0.194$^{***}$ \\ (0.009)} & \makecell{0.131$^{***}$ \\ (0.029)} & \makecell{0.067$^{***}$ \\ (0.014)} \\
\midrule
Skill FE & $-$ & $-$ & $-$ & x & $-$ & x \\
Target city FE & $-$ & $-$ & $-$ & $-$ & x & x \\
\midrule
Observations & 25,523 & 25,523 & 25,523 & 25,523 & 25,523 & 25,523 \\
$R^2$ & 0.003 & 0.18 & 0.181 & 0.235 & 0.229 & 0.291 \\
\bottomrule
\end{tabular}
}
\caption{Linear probability models excluding San Jose as origin. All variables standardized. Standard errors clustered by target city and skill.}\label{tab:horserace_no_sj}
\end{table}

\clearpage
\subsection*{Windows of locational opportunity: interaction test and robustness}

Table~\ref{tab:wlo_interaction} reports the pooled two-way fixed-effects interaction test underlying Section~\ref{sec:wlo}: a single regression of relatedness density on adoption lag, an emerging-skill indicator, and their interaction, with city and skill fixed effects and standard errors clustered two-way by city and skill. 

\begin{table}[h!]
\centering
\begin{tabular}{lc}
\toprule
 & Relatedness density \\
\midrule
Adoption lag & -0.0026 \\
 & (0.0002) \\
Lag $\times$ emerging & +0.0010 \\
 & (0.0005) \\
\midrule
Existing-skill gradient (std-$\beta$) & -0.111 \\
Emerging-skill gradient (std-$\beta$) & -0.070 \\
Observations & 76,463 \\
City-task cells (emerging / existing) & 10,331 / 66,132 \\
\bottomrule
\end{tabular}
\caption{(Pooled interaction test) the relatedness-density gradient on adoption lag for emerging vs.\ existing skills. City and skill fixed effects; standard errors clustered two-way by city and skill.}\label{tab:wlo_interaction}
\end{table}

\clearpage
\subsection*{PPML growth models}

Table~\ref{reg:PPML} presents Poisson Pseudo Maximum Likelihood estimates that combine intensive and extensive margins: how the number of users active in a skill in a city grows with relatedness density. The dependent variable is the number of users in city $c$ active in skill $\theta$ in the test period. Controls include base-period user counts (as a mean reversion term), city and skill fixed effects. Standard errors are clustered on city and skill.

\begin{table}[h!]
\centering
\resizebox{\linewidth}{!}{
\begin{tabular}{lccccccc}
   \toprule
    & \multicolumn{7}{c}{Number of users in skill at T+1}\\
                                       & (1) & (2) & (3) & (4) & (5) & (6) & (7)\\
   \midrule
   Relatedness Density at T & 8.677$^{***}$ & 8.201$^{***}$ & 2.529$^{***}$ & 2.243$^{***}$ & 1.866$^{***}$ & 1.852$^{***}$ & 1.344$^{***}$\\
                                       & (0.215) & (0.186) & (0.298) & (0.254) & (0.174) & (0.218) & (0.111)\\
   N users city C, skill Y, time T & & 0.009$^{***}$ & & & & & \\
                                       & & (0.001) & & & & & \\
   (Log) N users city C, skill Y, time T & & & 0.748$^{***}$ & 0.662$^{***}$ & 0.244$^{***}$ & 0.685$^{***}$ & 0.281$^{***}$\\
                                       & & & (0.032) & (0.027) & (0.021) & (0.025) & (0.020)\\
   Zero indicator for log(0) & & & $-$1.899$^{***}$ & $-$1.701$^{***}$ & $-$0.688$^{***}$ & $-$1.701$^{***}$ & $-$0.723$^{***}$\\
                                       & & & (0.071) & (0.066) & (0.047) & (0.058) & (0.040)\\
   (Log) FUA Population & & & & 0.216$^{***}$ & & 0.223$^{***}$ & \\
                                       & & & & (0.017) & & (0.017) & \\
   (Log) N Users, skill Y, time T & & & & 0.060 & 0.446$^{***}$ & & \\
                                       & & & & (0.046) & (0.042) & & \\
   \midrule
   Observations & 214,011 & 214,011 & 214,011 & 214,011 & 214,011 & 214,011 & 214,011\\
 \midrule
   City FE & & & & & Y & & Y\\
   Skill FE & & & & & & Y & Y\\
   \bottomrule
\end{tabular}
}
\caption{Poisson regression estimates of the number of users in a city active in a specific skill. Standard errors are clustered on city and skill.}\label{reg:PPML}
\end{table}

\clearpage

\subsection*{Robustness: Leave-one-out for the emergence regression}

The main-text emergence analysis takes the city as the unit of observation (Table~\ref{tab:emergence}). As a robustness check, we re-estimate that specification while dropping the top-contributing origin cities (San Jose and London) separately and jointly (Table~\ref{tab:emergence_robustness}).

\begin{table}[h!]
\centering
\renewcommand\cellalign{t}
\begin{threeparttable}
\begin{tabular}{lcccc}
\toprule
 & Full & Excl.\ San Jose & Excl.\ London & Excl.\ both \\
 & (1) & (2) & (3) & (4) \\
\midrule
\multicolumn{5}{l}{\emph{Panel A: OLS count $n_c$ (column 3 of Table~\ref{tab:emergence})}} \\
Log FUA population        & \makecell{0.042* \\ (0.018)} & \makecell{0.034* \\ (0.016)} & \makecell{0.030* \\ (0.014)} & \makecell{0.023 \\ (0.012)} \\
\addlinespace
Skill diversity (entropy) & \makecell{0.086** \\ (0.029)} & \makecell{0.070** \\ (0.024)} & \makecell{0.069** \\ (0.024)} & \makecell{0.053** \\ (0.017)} \\
\addlinespace
\multicolumn{5}{l}{\emph{Panel B: LPM, any origin (column 4 of Table~\ref{tab:emergence})}} \\
Log FUA population        & \makecell{0.012** \\ (0.004)} & \makecell{0.011** \\ (0.004)} & \makecell{0.010* \\ (0.004)} & \makecell{0.009* \\ (0.004)} \\
\addlinespace
Skill diversity (entropy) & \makecell{0.038*** \\ (0.010)} & \makecell{0.036*** \\ (0.010)} & \makecell{0.036*** \\ (0.010)} & \makecell{0.033** \\ (0.010)} \\
\midrule
Cities                    & 976 & 975 & 975 & 974 \\
Origins                   & 35  & 28  & 28  & 21 \\
\bottomrule
\end{tabular}
\footnotesize Significance: $*$ p $<$ 0.05, $**$ p $<$ 0.01, $***$ p $<$ 0.001. Robust standard errors.
\end{threeparttable}
\caption{Leave-one-out robustness for the city-level emergence regression. Columns (2) and (3) drop observations where San Jose or London (respectively) is the designated origin of any skill; column (4) drops both. Both coefficients remain positive and significant across all specifications, except population in the OLS count model when both hubs are dropped. \label{tab:emergence_robustness}}
\end{table}

\clearpage

\subsection*{City salary rankings and emerging tasks by city}
\rowcolors{1}{white}{white}
\begin{table}[h!]
\centering
\begin{tabular}{rlrr}
\toprule
Rank & City & Avg.\ Salary (\$) & Developers (FTE) \\
\midrule
1 & Beijing & \$131,627 & 2,794 \\
2 & Shenzhen & \$131,348 & 932 \\
3 & Hangzhou & \$131,179 & 679 \\
4 & Guangzhou & \$131,097 & 609 \\
5 & Shanghai & \$130,760 & 2,444 \\
6 & Karlsruhe & \$129,964 & 845 \\
7 & San Jose & \$129,932 & 13,110 \\
8 & Seoul & \$129,597 & 2,554 \\
9 & Cambridge (UK) & \$129,569 & 1,202 \\
10 & New Taipei [Taipei] & \$129,480 & 1,170 \\
11 & Grenoble & \$129,471 & 535 \\
12 & Saint Petersburg & \$129,429 & 2,891 \\
13 & Munich & \$129,408 & 4,594 \\
14 & Aachen / Heerlen & \$129,374 & 576 \\
15 & Tokyo & \$129,334 & 2,898 \\
16 & Moscow & \$129,131 & 6,542 \\
17 & Bengaluru & \$129,128 & 25,029 \\
18 & Nuremberg & \$129,112 & 756 \\
19 & Stuttgart & \$129,093 & 1,322 \\
20 & Boston & \$129,086 & 4,468 \\
\bottomrule
\end{tabular}
\caption{Top 20 cities by weighted average developer salary. Only cities with at least 500 full-time-equivalent developers are included. Salaries are computed as weighted averages of skill-level values from the Stack Overflow Developer Survey, where weights are local developer counts per skill.}\label{tab:cities_wages}
\end{table}

\begin{table}[h!]
\centering
\begin{tabular}{rlrr}
\toprule
Rank & City & Avg.\ Salary (\$) & Developers (FTE) \\
\midrule
1 & Abuja & \$124,955 & 517 \\
2 & Dhaka (outer) & \$125,414 & 2,139 \\
3 & Dhaka & \$125,515 & 6,739 \\
4 & Quezon City [Manila] & \$125,519 & 1,240 \\
5 & Accra & \$125,535 & 660 \\
6 & Lagos & \$125,776 & 3,188 \\
7 & Karachi & \$125,803 & 2,809 \\
8 & Kathmandu & \$125,846 & 2,067 \\
9 & Casablanca & \$125,900 & 611 \\
10 & Riga & \$125,949 & 578 \\
11 & Birmingham & \$126,011 & 709 \\
12 & Baku & \$126,034 & 621 \\
13 & Cape Town & \$126,074 & 1,488 \\
14 & Jakarta & \$126,165 & 1,632 \\
15 & Nottingham & \$126,168 & 618 \\
16 & Rajkot & \$126,199 & 843 \\
17 & Kampala & \$126,209 & 635 \\
18 & Tunis & \$126,211 & 1,152 \\
19 & Spring Hill & \$126,223 & 1,210 \\
20 & Edmonton & \$126,304 & 540 \\
\bottomrule
\end{tabular}
\caption{Bottom 20 cities by weighted average developer salary.}\label{tab:bottom_cities_wages}
\end{table}
\begin{table}[htbp]
\centering
\caption{The 35 emerging software skills and the city that originated each,
ordered by the number of skills a city originates. Skill descriptions are the
labels of the stochastic-block-model task communities.}
\label{tab:emerging_skills}
\footnotesize
\rowcolors{1}{gray!25}{white}
\begin{tabularx}{\textwidth}{@{}>{\raggedright\arraybackslash}X l@{}}
\toprule
Emerging skill & Originating city \\
\midrule
Build a multi-language Telegram bot using Python with Telethon, integrating Umbraco CMS for content management. & London \\
Develop a React Native application to visualize bioinformatics data using Firebase, Expo \& map navigation features. & London \\
Develop a neural network using TensorFlow to perform object detection on images, and visualize the results in TensorBoard. & London \\
Implement a currency conversion microservice with Spring Cloud, document API with Swagger, and monitor using Spring Boot Actuator. & London \\
Implement a service to fetch and store records with timestamps from InfluxDB using Retrofit2 and RxJava2. & London \\
Implement a web crawler using Scrapy to extract data for a Flutter app using provider for state management. & London \\
Integrate React components with Webpack, Axios, Enzyme, \& Fetch API, focusing on server-side rendering and model binding. & London \\
Configure MyPy in WebStorm for type hinting while typing JSX in a JetBrains IDE environment. & San Jose \\
Create a Python package supporting asyncio that uses aiohttp to display animated GIFs, compatible with Python 3.5-3.8. & San Jose \\
Create a cross-platform Android TV and mobile app with notifications, styled with Material Design, using NativeScript and Angular. & San Jose \\
Deploy a desktop application using Electron with PostgreSQL 9.4 backend, packaged with NSIS, to DigitalOcean VPS. & San Jose \\
Implement a multi-framework date picker feature within a SPA using AngularJS, React, and different UI libraries. & San Jose \\
Implement a splash screen with MVVMCross in a Xamarin.Forms app using getter/setter properties in Visual Studio Mac. & San Jose \\
Set up a Django project with Celery for background jobs, using Supervisord to manage worker processes spawned in VSCode. & San Jose \\
Create a CI/CD pipeline to build a Maven project, manage dependencies, and deploy to Kubernetes on AWS EKS using Ansible. & New York \\
Create a React form with Material-UI, use React Hook Form for validation, and submit data to SQL Server using IDisposable. & New York \\
Create a serverless workflow to process data using AWS Lambda, S3, and Step Functions, with autoscaling and monitoring setup. & New York \\
Develop a machine learning model to predict sentiment from text data, visualize results, and manage data workflow in Databricks. & New York \\
Implement an open-source TypeScript pathfinding algorithm in Cypress with GitHub Actions for CI, ensuring code quality with TSLint. & New York \\
Benchmark SaaS multi-tenant performance using JMeter, LoadRunner, and Spring WebFlux with HTTR and rvest for web scraping. & Seattle \\
Design a highly available, scalable streaming data pipeline using Apache Kafka and Confluent tools for real-time analytics. & Seattle \\
Implement an AG Grid in Angular 8 and populate it with data processed using Parallel.ForEach in Task Parallel Library. & Seattle \\
Integrate Azure AD with ASP.NET Core MVC for secure authentication using Azure Key Vault for secrets management. & Seattle \\
Build a Next.js application on a monorepo, with Twitter integration, Algolia search, and hosted on Vercel using Firebase for authentication. & Bengaluru \\
Set up a systemd service to run a Java daemon with audio processing using ALSA and PyAudio, and enforce code style with ESLint and Prettier. & Bengaluru \\
Develop a React application with form validation and state management, integrating GraphQL and automated testing. & Washington D.C. \\
Set up a webpack development environment to bundle a Babel-transpiled app querying SPARQL endpoints with DBpedia. & Washington D.C. \\
Implement a data pipeline using Apache Flink on Cloud Foundry, managed by Spring Cloud Dataflow, in a JHipster application. & Austin \\
Set up a Jenkins pipeline to run unit tests using various frameworks and mock libraries across different JavaScript projects. & Boston \\
Upgrade an Angular 6 app to use TypeScript 2.0, RxJS 6, and Angular Reactive Forms with server-sent events. & Buenos Aires \\
Build a responsive website with Gatsby, styled with Styled Components, and deploy on Netlify with a Strapi CMS backend. & Copenhagen \\
Create a multi-cloud analytics platform using Terraform to deploy Elasticsearch on AWS, Apache Spark on GCP, and handle streaming data. & Denver \\
Develop a Google Glass app with Android Studio using Android SDK tools and Gradle for the build process. & Leeds \\
Integrate Azure AD with an iOS app to authenticate using MSAL and manage notifications with NSNotificationCenter. & Portland \\
Implement a SwiftUI list fetching data using Alamofire with refreshable content on iOS 14 and JSON parsing. & Turku \\
\bottomrule
\end{tabularx}
\label{tab:emerging_skills}
\end{table}

\begin{landscape} % Begin landscape orientation
\newcolumntype{P}[1]{>{\raggedright\arraybackslash}p{#1}}
\noindent
\begin{table}[h!]
\centering
\footnotesize
\rowcolors{3}{gray!25}{white}
\begin{tabular}{P{3cm}|P{3.4cm}|P{3.4cm}|P{3.4cm}|P{3.4cm}|P{3.4cm}}
\toprule
City & Skill 1 & Skill 2 & Skill 3 & Skill 4 & Skill 5 \\
\midrule
San Jose & Implement immutable collections in Scala using traits, compare their equality, and manage dependencies with SBT and Conda. & Create a JSFiddle snippet to demonstrate a MooTools animation, and analyze text files using grep, cat, xetails cut. & Design a bootloader using NASM for x86\_16 CPUs with optimized assembly routines considering calling conventions and CPU cache utilization. & Optimize Emacs org-mode loading large files with custom key bindings and efficient garbage collection handling. & Explain how passing by reference differs from passing by value using anonymous functions and mutable types in delegates and protocols. \\
\midrule
Berlin & Analyze the effects of categorical data on mixed models' confidence intervals using various Scala libraries for type inference. & Implement a data pipeline using Apache Flink on Cloud Foundry, managed by Spring Cloud Dataflow, in a JHipster application. & Create a recommendation engine using polymorphic associations in ActiveRecord with GTK and Cairo for the interface. & Create a Joomla extension to enhance SEO with optimized titles and meta tags for blog posts. & Explain how passing by reference differs from passing by value using anonymous functions and mutable types in delegates and protocols. \\
\midrule
Bengaluru & Implement a secure Hadoop-based data pipeline using Netty for non-blocking I/O and Kerberos authentication. & Create a JSFiddle snippet to demonstrate a MooTools animation, and analyze text files using grep, cat, xetails cut. & Design a highly available, scalable streaming data pipeline using Apache Kafka and Confluent tools for real-time analytics. & Develop a Java EE web application using Struts2, EJB 3.1, and JSP with deployment on WildFly server. & Automate LinkedIn profile updates using Camunda BPMN workflow triggered by Alfresco Share events via Guzzle HTTP requests. \\
\midrule
Beijing & Implement a function to manage vehicle identification numbers using STL containers like unordered\_map, vector, and string. & Implement a data pipeline using Apache Flink on Cloud Foundry, managed by Spring Cloud Dataflow, in a JHipster application. & Implement a P2P WebRTC live streaming platform with adaptive bitrate using HLS and RTMP protocols. & Develop a multitasking application to edit RTF documents, with autosave, load functions, and inter-process communication via pipes. & Create a countdown timer on a MediaWiki page to track an upcoming Wikipedia event using the Wikipedia API. \\
\midrule
Moscow & Implement a numerically solved differential equation in Simulink and replicate the solution using Wolfram Mathematica. & Create a cross-platform C++ financial application using Boost libraries, with stock data fetched from Yahoo Finance. & Migrate a PyQt4 application to PySide2 or PyQt5 using Qt Designer for UI enhancement while managing background processes with QThread. & Integrate Laravel Passport authentication with a Spring and Apache Camel based JMS messaging system. & Create a 64-bit union for handling different pointer types and endianness conversions for debugging memory dumps. \\
\midrule
Vienna & Create a Joomla extension to enhance SEO with optimized titles and meta tags for blog posts. & Create a JSFiddle snippet to demonstrate a MooTools animation, and analyze text files using grep, cat, xetails cut. & Integrate Azure AD with an iOS app to authenticate using MSAL and manage notifications with NSNotificationCenter. & Create a .NET Standard application with a Tkinter GUI, packaged with cx\_Freeze, and distributed via ClickOnce. & Develop an AI to optimize webcam video capture quality using eigenvalue-based PCA and genetic algorithm smoothing with RTSP and H.264. \\
\midrule
Rio de Janeiro & Implement user authentication in DotNetNuke using ASP.NET membership to prevent deadlocks with PostgreSQL 9.3 and channels. & Create a sliding color box animation with adjustable opacity and width compatibility for various Internet Explorer versions. & Configure Tomcat8 to output application logs using Log4j2, and integrate with syslog for centralized logging. & Develop a blockchain smart contract for secure data storage using Hyperledger Fabric and integrate it with an IBM DB2 database. & Create a kineticjs canvas to simulate a rotating object at a given rpm based on specifications. \\
\bottomrule
\end{tabular}
\caption{The highest RCA skills for a select group of cities.}\label{tab:city_rca_skills}
\end{table}
\end{landscape}

%% FOR PRESENTATION TABLES
% \newpage
% \begin{table}[h!]
% \centering
% \begin{tabular}{rlrr}
% \toprule
% Rank & City & Avg.\ Salary (\$) & Developers (FTE) \\
% \midrule
% 1 & Beijing & \$131,627 & 2,794 \\
% 2 & Shenzhen & \$131,348 & 932 \\
% 3 & Hangzhou & \$131,179 & 679 \\
% 4 & Guangzhou & \$131,097 & 609 \\
% 5 & Shanghai & \$130,760 & 2,444 \\
% 6 & Karlsruhe & \$129,964 & 845 \\
% 7 & San Jose & \$129,932 & 13,110 \\
% 8 & Seoul & \$129,597 & 2,554 \\
% 9 & Cambridge (UK) & \$129,569 & 1,202 \\
% 10 & New Taipei [Taipei] & \$129,480 & 1,170 \\
% \bottomrule
% \end{tabular}
% \caption{Top 20 cities by weighted average developer salary. Only cities with at least 500 full-time-equivalent developers are included. Salaries are computed as weighted averages of skill-level values from the Stack Overflow Developer Survey, where weights are local developer counts per skill.}
% \end{table}

% \begin{table}[h!]
% \centering
% \begin{tabular}{rlrr}
% \toprule
% Rank & City & Avg.\ Salary (\$) & Developers (FTE) \\
% \midrule
% 1 & Abuja & \$124,955 & 517 \\
% 2 & Dhaka (outer) & \$125,414 & 2,139 \\
% 3 & Dhaka & \$125,515 & 6,739 \\
% 4 & Quezon City [Manila] & \$125,519 & 1,240 \\
% 5 & Accra & \$125,535 & 660 \\
% 6 & Lagos & \$125,776 & 3,188 \\
% 7 & Karachi & \$125,803 & 2,809 \\
% 8 & Kathmandu & \$125,846 & 2,067 \\
% 9 & Casablanca & \$125,900 & 611 \\
% 10 & Riga & \$125,949 & 578 \\
% \bottomrule
% \end{tabular}
% \caption{Bottom 20 cities by weighted average developer salary.}
% \end{table}

\end{document}